\documentclass[%
reprint,
 amsmath,amssymb,
 aps,
floatfix
]{revtex4-2}

\usepackage{graphicx}
\usepackage{dcolumn}
\usepackage{bm}

\usepackage[utf8]{inputenc}
\DeclareUnicodeCharacter{2212}{-}

\usepackage{mathtools}

\usepackage{afterpage}

\usepackage{relsize}

\newcommand{\reals}{\mathbb{R}}
\newcommand{\integers}{\mathbb{Z}}
\newcommand{\nats}{\mathbb{N}}

\usepackage{amsthm}
\newtheorem*{definition}{Definition}
\newtheorem*{theorem}{Theorem}

\usepackage{tikz}
\usetikzlibrary{external}
\begin{document}

\preprint{}

\title{Quantitative analysis of phase transitions in\\two-dimensional XY models using persistent homology}

\author{Nicholas Sale}
\email{nicholas.j.sale@gmail.com}
\author{Jeffrey Giansiracusa}
  \altaffiliation[Also ]{Department of Mathematical Sciences, Durham University, Upper Mountjoy Campus, Durham, DH1 3LE, UK}
 \author{Biagio Lucini}
  \altaffiliation[Also ]{Swansea Academy of Advanced Computing, Swansea University, Bay Campus, SA1 8EN, Swansea, Wales, UK}
\affiliation{
 Department of Mathematics, Swansea University, Bay Campus, SA1 8EN, Swansea, Wales, UK
}

\date{\today}

\begin{abstract}
We use persistent homology and persistence images as an observable of three different variants of the two-dimensional XY model in order to identify and study their phase transitions. We examine models with the classical XY action, a topological lattice action, and an action with an additional nematic term. In particular, we introduce a new way of computing the persistent homology of lattice spin model configurations and, by considering the fluctuations in the output of logistic regression and k-nearest neighbours models trained on persistence images, we develop a methodology to extract estimates of the critical temperature and the critical exponent of the correlation length. We put particular emphasis on finite-size scaling behaviour and producing estimates with quantifiable error. For each model we successfully identify its phase transition(s) and are able to get an accurate determination of the critical temperatures and critical exponents of the correlation length.
\end{abstract}

\maketitle

\parskip=10pt

\section{\label{sec:intro}Introduction}
There is an emerging body of work exploring the use of machine learning and other data analysis methods to detect and classify phase transitions in statistical physics systems. An incomplete list of references includes \cite{Carrasquilla2016MachineLP, Wang2016DiscoveringPT, Akinori2017DetectionOP, Nieuwenburg2017LearningPT, Wetzel2017MachineLO, Greitemann2018ProbingHS, Canabarro2019UnveilingPT,Alexandrou:2019hgt,Rzadkowski:2019jsz,Bachtis:2020dmf,Bachtis:2020ajb,Bachtis:2020fly,Bachtis:2021xoh,Bachtis:2021eww}. One of the motivations of this approach is to develop methodologies which require minimal \textit{a priori} knowledge about the systems in question. The hope then is that these data-centric methods will be able to offer new insights into those models at the forefront of physics which seem to defy analytical methods \cite{GIANNETTI2019114639}. Much of the work in this area makes use of neural network models which, while unparalleled in machine learning tasks, are generally hard to interpret. But recently, among other geometric and topological approaches \cite{Rodriguez-Nieva2019, mendessantos2020unsupervised, PhysRevE.100.032414, PhysRevB.99.085138}, there has been an interest in using persistent homology, a tool from the new field of topological data analysis (TDA), to produce interpretable features which are inherently sensitive to topological objects. These can then be compared in their own right, or fed into a machine learning model \cite{PhysRevE.98.012318, Hirakida2018PersistentHA, tran2020topological, olsthoorn2020finding, cole2020quantitative, Donato2016PersistentHA}.

There are at least two paradigms for using persistent homology to study phase transitions of a given statistical physics model. The first can be called \textit{persistent homology in configuration/data space}, where the topology of the high-dimensional space of model configurations is probed from samples. This approach is based on the \textit{topology hypothesis} for the origin of phase transitions \cite{PhysRevLett.79.4361, Kastner2008} and is the approach used in \cite{Donato2016PersistentHA}. The idea here is that a thermodynamic phase transition necessarily coincides with a change in the topology of the energy level set, although such a change does not turn out to be a sufficient condition \cite{PhysRevLett.92.060601}. In the present work however, we shall make use of a newer paradigm, investigated also in \cite{PhysRevE.98.012318, Hirakida2018PersistentHA, tran2020topological, olsthoorn2020finding, cole2020quantitative}, which we call \textit{persistent homology as an observable}. Given a sampled configuration of a model, we construct a sequence of geometric complexes based on that configuration. This sequence of topological spaces is known as a filtration. Applying persistent homology to the filtration yields a collection of points called a persistence diagram, which represents this configuration. We can think of this process as a means to reduce the degrees of freedom of the model and produce nonlinear summaries of configurations. Statistics of these persistence diagrams are then analysed as the system undergoes a phase transition. Previous works have focused on identifying the different phases in various models in a mostly qualitative manner. While \cite{cole2020quantitative} makes some steps towards obtaining quantitative measurements of the multiscale structure of the Ising model at criticality, a framework for using persistent homology observables to make rigorous numerical estimates of critical temperatures and exponents with quantified error has not yet been explored in the literature.

While the existing works on the \textit{persistent homology as an observable} paradigm share the same underlying idea, the approaches seen so far have differed significantly, both in how filtrations have been constructed, and how the resulting persistence diagrams have been analysed. Tran, Chen, and Hasegawa investigated phase transitions in the 2D XY model, 1D transverse-field Ising and 1D Bose-Hubbard models \cite{tran2020topological}. They computed the Vietoris-Rips persistence of point clouds of lattice sites with inter-point distances given by a linear combination of the Euclidean distance in the lattice and the difference in the spins. They show that clustering configurations based on the the Persistence Fisher kernel \cite{NIPS2018_8205}, persistence entropy and the second moment of persistence of the $H_1$ diagrams identifies the different phases. They demonstrate that increasing the lattice size produces sharper estimates of the critical temperature. This approach is extended to the XXZ model on a pyrochlore lattice by Olsthoorn, Hellsvik, and Balatsky, approximately separating the six different phases of the model \cite{olsthoorn2020finding}. 

Cole, Loges, and Shiu apply a different methodology to the previous works. Looking at the 2D Ising, square-ice, XY and fully-frustrated XY models, they introduce general constructions of filtrations for configurations of discrete-valued and circle-valued spin models \cite{cole2020quantitative}. In particular, configurations of circle-valued models are given a sublevel set filtration of the map $f: \Lambda \rightarrow (-\pi, \pi]$ which assigns each site $i \in \Lambda$ in the lattice a parameterisation of its spin $f(i) \in (-\pi, \pi]$. This filtration yields cubical subcomplexes of the lattice. They make use of persistence images \cite{persistenceImages} to vectorise persistence diagrams, allowing the application of a logistic regression model to separate the phases. They relate some quantitative aspects of the persistence diagrams to the estimation of critical exponents in the case of the Ising model. For discrete models they construct $\alpha$-complexes on subsets of the lattice sites with the same spin. This is similar to the approach used by Hirakida et al. in \cite{Hirakida2018PersistentHA} who look at the effective Polyakov line model. 

Comparing the approach in \cite{tran2020topological} and \cite{olsthoorn2020finding} to that in \cite{cole2020quantitative} makes it clear that there is a significant degree of choice in picking the filtration used to compute the persistent homology of a given lattice configuration. We will demonstrate that this choice is an important factor in determining what information about phase transitions one can derive from the persistence. In particular, we investigate an XY model with a nematic interaction term and find that using two different filtrations is required to detect and analyse the two different phase transitions undergone by the system.

Our main contributions are as follows:
\begin{itemize}
    \item We introduce a new class of filtrations on lattice spin systems which, while general, allow persistent homology to easily detect topological defects.
    \item Extending the approach of using logistic regression on persistence images introduced in \cite{cole2020quantitative}, we investigate the applicability of finite-size scaling analysis. In particular, we apply the standard statistical tools of histogram reweighting and bootstrapping to obtain estimates of the critical temperature and the critical exponent of the correlation length with quantified error.
    \item Finding inadequacies with using logistic regression for precise estimates of the critical temperature, we introduce a non-parametric method using k-nearest neighbour classification as a tool to estimate the critical temperature of phase transitions from persistence images. This yields improved results.
    \item We consider a model with both an Ising-type and  Berezinskii-Kosterlitz-Thouless (BKT) transition (the Nematic XY model) and find that two different filtrations are required to capture the two transitions. Each filtration sees one transition, but neither is able to capture information about both transitions. We take this as evidence that the technique is not applicable entirely unsupervised; rather, care must be taken to design a filtration tuned for the problem.
\end{itemize}

The rest of the paper is organised as follows. In Section \ref{sec:method} we give a brief review of the techniques we use, covering persistent homology, supervised classification, finite-size scaling analysis, histogram reweighting and bootstrapping. At the end we detail the steps of the data generation and analysis pipeline. In Section \ref{sec:analysis} we look at the three models under consideration. In each case we give a brief review of the model and its phase transition(s) before discussing the analysis and results using logistic regression and then k-nearest neighbours. In Section \ref{sec:conclusions} we discuss our findings and identify potential directions for future work. The appendices contain more detailed reviews of some of the tools we use as well as the argument demonstrating the stability of the persistence diagrams obtained using our filtrations.

\section{Method}
\label{sec:method}

\subsection{Background on Persistent Homology and Persistence Images}
Persistent homology is a computational topology tool introduced in its modern form in \cite{Edelsbrunner2002TopologicalPA} and popularised in \cite{Carlsson2009TopologyAD}. It is one of the main tools of the emerging field of Topological Data Analysis. We shall give a brief overview here, but for a more complete review of persistent homology useful references are \cite{carlsson2020persistent, ph_survey_edels_harer, otter_roadmap, Ghrist2007BarcodesTP}. 

Given a topological space, such as a manifold or a simplicial/cubical complex, homology can intuitively be thought of as an algebraic way of describing the 'holes' in the space. In particular, the spaces we consider will be cubical complexes. A very brief technical introduction to cubical complexes and their homology can be found in Appendix \ref{appendix:cubical}. But in general terms, given a cubical complex $C$, its $k$\textsuperscript{th} cubical homology $H_k(C)$ is a vector space which has a basis in $1$-$1$ correspondence with the $k$-dimensional holes in $C$. Moreover, given a map of cubical complexes $f : C \rightarrow C^\prime$, we obtain induced linear maps $f_k: H_k(C) \rightarrow H_k(C^\prime)$. The rank of $f_k$ tells us how many of the holes survived after being mapped into $C^\prime$ i.e. how many \textit{persisted}. Given some data $D$, the idea of (cubical) persistence then is to construct a sequence
$$F_1(D) \rightarrow F_2(D) \rightarrow \ldots \rightarrow F_N(D)$$
of cubical complexes called a filtration using the data. Typically the $F_i(D)$ are each subcomplexes of the final complex $F_N(D)$; for each cell we specify the index $i$ at which it appears, and then $F_i$ is the subcomplex consisting of all cells that have appeared at or before $i$.  The maps $F_i(D) \to F_{i+1}(D)$ are simply the inclusions. 

We then apply homology to obtain a sequence of linear maps
$$H_k(F_1(D)) \rightarrow H_k(F_2(D)) \rightarrow \ldots \rightarrow H_k(F_N(D)).$$
Using the ranks of these maps we can track the birth of new holes, their persistence through the filtration, and their deaths. We summarise this information as a multi-set called a persistence diagram $PH_k(F(D)) \subset \{ (a,b) \in \reals^2 \mid a \leq b \}$ which contains a pair $(b,d)$ every time a hole is born in $F_b(D)$ and dies in $F_d(D)$. We say that a feature is born at $b$, dies at $d$ and that its persistence is $d-b$. This can also be represented as a barcode (a multi-set of intervals $[b, d)$). There are a few ways to define distances between persistence diagrams, but those which are most commonly used are the bottleneck and Wasserstein distances. For many typical choices of filtration a small change in the input data $D$ leads to only a small change in the persistence diagram $PH_K(F(D))$ as measured by these distances. This property of persistent homology is known as stability, and makes persistence a useful tool for dealing with real-world, noisy data. 

In the \textit{persistent homology in configuration space} paradigm, $D$ is the entire collection of sampled configurations, and we obtain a single persistence diagram. However, when we use persistent homology as an observable, $D$ is a single configuration of the model we are studying. We therefore obtain a persistence diagram for each sampled configuration and we can consider statistics computed from these diagrams. Unfortunately persistence diagrams in their raw form as multi-sets do not lend themselves to computing the typical statistics of interest such as means and variances. While there has been work developing notions of these quantities as Frech\'et means and variances \cite{Turner2014FrchetMF}, we shall instead prefer to work with a vector representation of the diagrams known as persistence images \cite{persistenceImages} which preserve stability. 

\begin{figure}[ht]
    \centering
    \scalebox{0.36}{\includegraphics{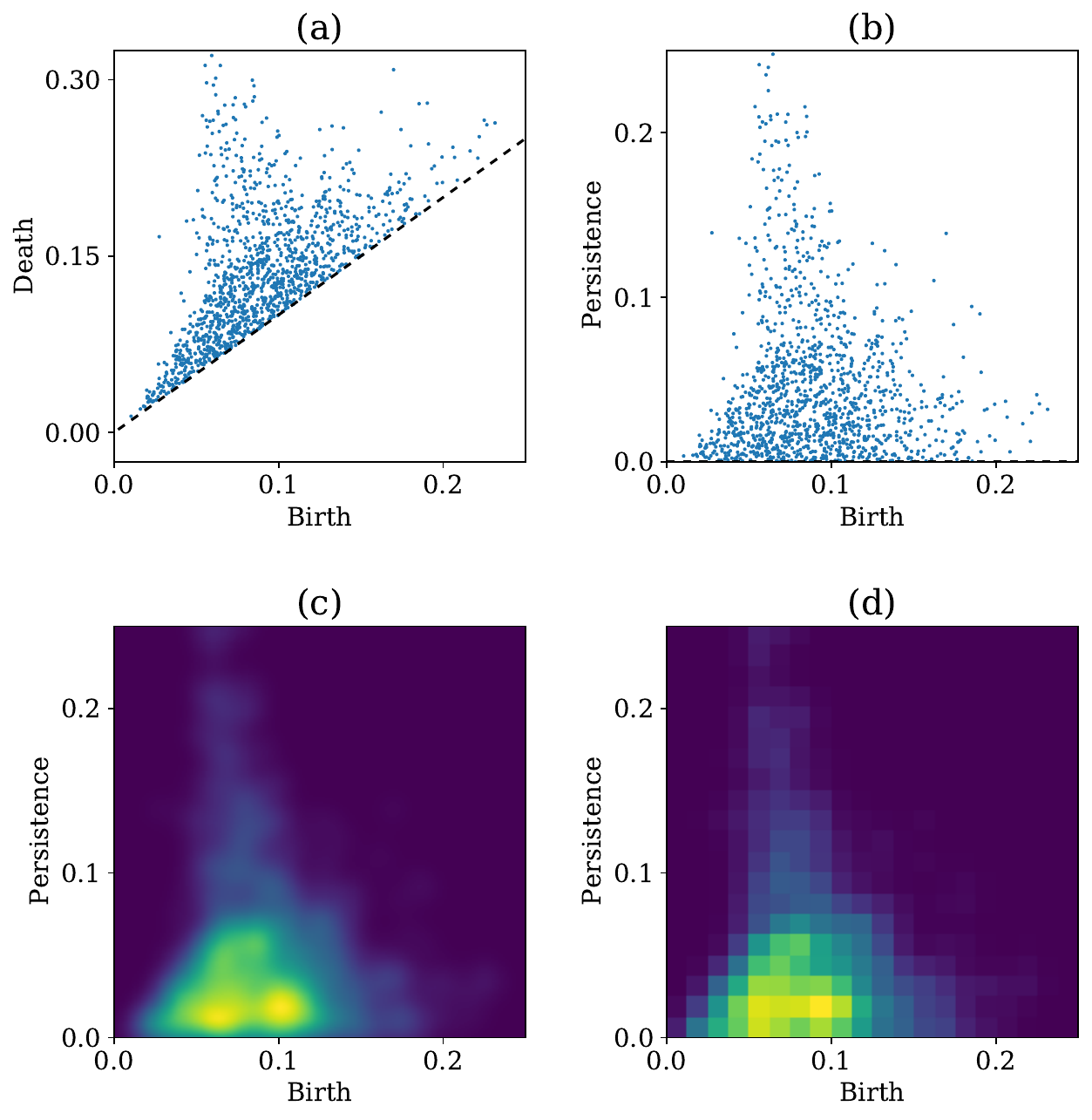}}
    \caption{An illustration of how the persistence image is obtained from a persistence diagram (a). It is first transformed into birth-persistence coordinates (b), then the persistence surface (c) is computed before discretisation, yielding the persistence image (d).}
    \label{fig:pi}
\end{figure}

Let $g_{a,b} : \reals^2 \rightarrow \reals$ denote a 2D Gaussian of standard deviation $\sigma$ centered at $(a,b)$: $$g_{a,b}(x,y) = \frac{1}{2\pi\sigma^2}exp\bigg[-\frac{(x-a)^2 + (y-b)^2}{2\sigma^2}\bigg].$$ Given a persistence diagram $PH_k = \{ (b_i, d_i) \}_{i \in I}$, its persistence surface is the function $\rho_k : \reals^2 \rightarrow \reals$ obtained by translating each point $(b,d) \in PH_k$ into birth-persistence coordinates $(b, d-b)$, then placing Gaussians with variance $\sigma^2$ on them, weighted by the persistence of the point:
$$\rho_k(x,y) = \sum_{(b,d)\in PH_k}(d-b) \, g_{b,d-b}(x,y).$$
The persistence image $PI_k$ is obtained by discretizing a rectangular region of the domain of $\rho_k$ into a collection of $n_I \times n_I$ pixels $p_i$ and integrating $\rho_k$ within each:
$$PI_k^i = \iint_{p_i} \rho_k(x,y)dxdy.$$
In this way we obtain a $(n_I)^2$-dimensional vector representing our persistence diagram. See Figure \ref{fig:pi} for an example. So long as we choose the same $\sigma$ and discretization for each diagram, we can compute averages and variances component-wise. As observed in \cite{DBLP:journals/jocg/DivolC19}, if we are sampling data from some distribution and the expected persistence diagram has a density with respect to the Lebesgue measure on $\{ (a,b) \in \reals^2 \mid a \leq b \}$, then the average of the persistence images can be thought of as an estimator for this density, multiplied by an additional weighting equal to the persistence. Besides emphasising high-persistence points, the linear weighting by the persistence ensures the stability of the persistence image. Finally we note that, as discussed in \cite{persistenceImages}, machine learning models trained on persistence images are generally insensitive to the resolution and variance parameters $n_I$ and $\sigma$. Therefore in this work, we shall fix the parameters with a resolution of $30 \times 30$ and $\sigma$ equal to $10\%$ of a pixel. However, as a check we also performed one of the later experiments with a $15 \times 15$ resolution, finding no significant change in the results or estimated errors.

\subsection{Filtrations}
\label{sec:filtration}

In this work we will be working with finite two-dimensional lattices with a circle valued spin at each lattice site. To apply persistent homology we must choose how to define a filtration for a given configuration $\boldsymbol{\theta} = \{\theta_i\}$, where $\theta_i$ represents the spins as angles. Our idea is to filter the square tiling of the plane corresponding to the lattice $\Lambda$ according to the differences in neighbouring spins. For each cell in this cubical complex, we will specify a time at which it appears, and then $F_t$ is the subcomplex of the plane consisting of all cells that have appeared by time $t$.  Denote the smallest angle between spins $\theta_i$ and $\theta_j$ by $d_{ij}$. This can also be seen as the length of the shortest arc between $\theta_i$ and $\theta_j$ on the unit circle. Then taking the lattice as a 2-dimensional cubical complex, we introduce each vertex $i$ at time $0$, each edge $\langle ij \rangle$ at time $\frac{1}{2\pi}{d_{ij}}$, and each plaquette $\square$ at time $\text{max}_{i,j\in\square}\,\frac{1}{2\pi}{d_{ij}}$. We will call this the angle difference filtration. We will also introduce another similar filtration to use with the Nematic XY model in Section \ref{sec:nematic}. This will instead use a nematic angle difference $d_{ij}^n$ which denotes the smallest angle between the spins $\theta_i$ and $\theta_j$ considered as directionless rods. We can think of this as the length of the shortest arc connecting the head of one spin to either the head or tail of the other spin. That is $d_{ij}^n = \min(d_{ij}, \pi - d_{ij})$. We will call this the nematic angle difference filtration. 

\begin{figure*}[ht]
    \centering
    \scalebox{0.75}{\includegraphics{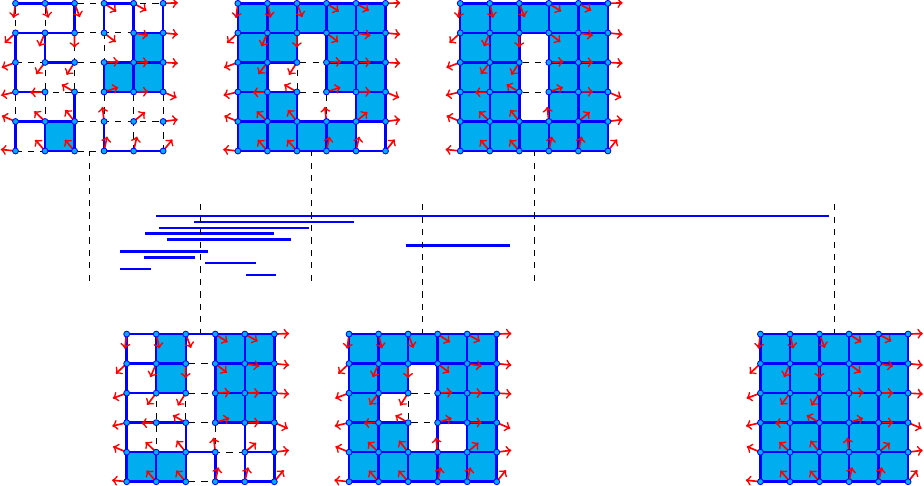}}
    \caption{An illustration of the angle difference filtration for a configuration of the XY model with an antivortex. The filtration parameter increases from left to right and the state of the filtration is shown at 6 different stages. On the left-hand side only those neighbouring spins which don't differ too much are connected by edges and plaquettes. As we move towards the right, more and more edges are introduced between more disparate spins. Note the correspondence between the bars and the holes in the filtration. For example, the longest bar corresponds to the hole around the antivortex in the centre of the configuration. This hole is formed early on as the spins far from the centre vary slowly, but survives until the central plaquette is added to the filtration.}
    \label{fig:barcode}
\end{figure*}

The intuition behind these filtrations originally came from considering the 2D XY model. Regions of the lattice where spins vary slowly will be introduced in the angle difference filtration early, while regions containing rapidly varying spins, such as at the centre of vortices, will enter the filtration later. We should expect then, at least at low temperatures, that each vortex will be manifested as a hole in the filtered lattice which is formed early on in the filtration, and which only gets filled in much later: i.e. a persistent $H_1$ class. Figure \ref{fig:barcode} shows an example of this. However we will see that this kind of filtration can capture other structure such as spin waves, or half-vortices and domain walls when we look at the Nematic XY model. Moreover, compared to the point cloud filtrations used in \cite{tran2020topological, olsthoorn2020finding} this class of filtrations has the computational benefit that edges are only introduced between neighbouring lattice sites and only elementary cubes up to dimension 2 are included, greatly speeding up the computation of persistent homology. In this case the filtrations consist of subcomplexes of the plane, so contain cubes of dimension at most 2 anyway. But note that for models on higher dimensional lattices, including cubes of higher dimension in the filtration would not have any effect on $H_1$ which is the only homological degree we use in our analysis. As discussed in Appendix \ref{appendix:stability}, the persistence diagrams obtained using these filtrations are stable with respect to small perturbations to the spins, in contrast to the sublevel set filtration used in \cite{cole2020quantitative}.

\subsection{Logistic Regression}
\label{sec:logistic_regression}
Following the approach introduced in \cite{cole2020quantitative}, we will train a logistic regression model to map the persistence images obtained from configurations onto phases. Recall that logistic regression is a generalised linear model which models a binary dependent variable $y(\mathbf{x}) \in \{0, 1\}$. For input $\mathbf{x} \in \reals^N$, a logistic regression model is parameterised by a weight vector $\mathbf{w} = (w_1, \dots, w_N)^T \in \reals^N$ and intercept $b \in \reals$. Its output is a logistic function
$$p_{\mathbf{w},b}(\mathbf{x}) = \frac{1}{1+e^{\mathbf{x}^T \mathbf{w} + b}} \in (0, 1)$$
which can be interpreted as the probability that $y(\mathbf{x}) = 1$, with $1 - p(\mathbf{x})$ giving the probability that $y(\mathbf{x}) = 0$. Given training data $\{(\mathbf{x}_i, y_i)\}$, the weights $\mathbf{w}$ and intercept $b$ are learnt by minimising a cross-entropy loss function
\begin{multline*}
    J(\mathbf{w}, b) = \\- \sum_i \big[ y_i log(p_{\mathbf{w},b}(\mathbf{x}_i)) + (1 - y_i) log(1 - p_{\mathbf{w},b}(\mathbf{x}_i)) \big]\\ + \frac{1}{C} (\mathbf{w}^T \mathbf{w} + b^2).
\end{multline*}
The first term penalises misclassifications with the penalty increasing as the confidence in the incorrect classification increases. The second term implements $\ell_2$ regularisation, reducing overfitting by preventing the weights from becoming too large, where $C$ is a hyper-parameter controlling the amount of regularisation. 

In our case, $\mathbf{x}$ will be a persistence image, $y(\mathbf{x}) = 0$ will indicate the low-temperature phase, and $y(\mathbf{x}) = 1$ will indicate the high-temperature phase. As in \cite{cole2020quantitative} we will train the model using data drawn in the low and high temperature phases. However, since we are interested in making a precise estimate of the critical temperature, we will use data closer to the critical region. After successful training, the weights will indicate features in the persistence image characteristic of each phase. Weights $w_j < 0$ will indicate features of the low-temperature phase, and weights $w_j > 0$ will indicate features of the high-temperature phase. In the intermediate range of temperatures where there is no training data, the logistic regression model will output an estimated classification $O_{LR} \in \{0,1\}$ depending on whether $p$ is less than or greater than $0.5$. Notice that we clamp the output to $0$ or $1$ rather than using the direct output of the logistic function. We find that this leads to better finite-size scaling behaviour later on. We may then treat $\langle O_{LR} \rangle$ as a phase indicator, if not a true (dis)order parameter. In this work we shall be interested in the distribution $O_{LR}$ at different temperatures and different lattice sizes.

We note that training the logistic regression model directly on raw configurations is ineffective due to the highly nonlinear nature of the system.

\subsection{k-Nearest Neighbours Classification}
\label{sec:knn}

We will also make use of k-nearest neighbours (k-NN) classification to map persistence images onto phases. This is a non-parametric model which models a categorical dependent variable $y(\mathbf{x}) \in \nats$, where $\mathbf{x} \in \reals^N$. The behaviour of the model is determined by the training data $\{(\mathbf{x}_i, y_i)\}$ and a choice of the hyper-parameter $k \in \nats$. Given new input $\mathbf{x}$, it finds the $k$ indices $i^1_\mathbf{x}, \ldots, i^k_\mathbf{x}$ which minimise the Euclidean distance $\vert\vert \mathbf{x} - \mathbf{x}_i \vert\vert_2$. It then outputs the most common label among the $y_{i^1_\mathbf{x}} ,\ldots, y_{i^k_\mathbf{x}}$. 

As in the case of logistic regression, $\mathbf{x}$ will be a persistence image, $y(\mathbf{x}) = 0$ will indicate the low-temperature phase, and $y(\mathbf{x}) = 1$ will indicate the high-temperature phase. We will train the model using data drawn in the low and high temperature phases close to the critical region. In the intermediate range of temperatures where there is no training data, the k-NN model will output an estimated classification $O_{kNN} \in \{0,1\}$. We may then treat $\langle O_{kNN} \rangle$ as a phase indicator. 

We note that training the classifier directly on raw configurations is not computationally feasible; doing so would require a vastly larger number of samples to sufficiently fill out the configuration space and the computational cost of the kNN method would consequently grow too large.  The mapping from configurations to persistence images concentrates the distribution near a low-dimensional subspace, and hence kNN becomes effective with far fewer samples.

\subsection{Finite-Size Scaling Analysis}
\label{sec:finite_size_scaling}
A typical approach to extracting the critical temperature and critical exponents of continuous phase transitions in spin systems is a finite-size scaling analysis of quantities such as the magnetic susceptibility
$$\chi(T) = \frac{L^2}{T} \big[\langle \vert M \vert^2 \rangle_T - \vert\langle M \rangle\vert^2_T \big]$$
which diverges at the critical temperature in the thermodynamic limit, where $M = L^{-2} \sum_i (\cos \theta_i ,\, \sin \theta_i )$ is the magnetisation vector. On a finite lattice of length $L$ this quantity will remain analytic, instead displaying a pronounced peak at a pseudo-critical temperature somewhere above or below the true critical temperature $T_c$. As $L \rightarrow \infty$ this peak grows taller and moves closer towards $T_c$. For a second-order phase transition, like that in the Ising model, the way in which the susceptibility scales close to $T_c$ can be described by the form
\begin{equation}
\label{eqn:susceptibility_fss_secondorder}
    \chi(L, t) = L^{\gamma / \nu} \, \hat{\chi}(L^{1 / \nu} \, t)
\end{equation}
where $\hat{\chi}$ is a dimensionless function, $t = \frac{T - T_c}{T_c}$ is the reduced temperature, and $\gamma$ and $\nu$ are the critical exponents for the susceptibility and correlation length respectively. For a BKT transition, like that in the 2D XY model, it scales approximately according to
\begin{equation}
\label{eqn:susceptibility_fss_bkt}
    \chi(L, t) \approx L^{\gamma / \nu} \, \hat{\chi}(L \exp(-bt^{-\nu}))
\end{equation}
where we have ignored some small logarithmic corrections. By simulating close to the phase transition on different lattice sizes $L$ we can extract the heights and locations of the different peaks then fit these to Equation \ref{eqn:susceptibility_fss_secondorder} or Equation \ref{eqn:susceptibility_fss_bkt} as appropriate to estimate $T_c$, $\gamma$ and $\nu$. Note that the logarithmic corrections we ignored in the BKT case mean that this method is not typically used for high precision studies, where approaches based on the spin stiffness are more common. 

Analogously, we might expect the persistent homology of a configuration to demonstrate large variations at criticality. We quantify this by looking at the fluctuations in the output $O_{LR}$ and $O_{kNN}$ of the trained logistic regression and k-NN models, measuring the variance
\begin{align}
\begin{split}
        \chi_{LR}(T) & = \langle O_{LR}^2 \rangle_T - \langle O_{LR} \rangle^2_T\\ & = \langle O_{LR} \,\, \rangle_T (1 - \langle O_{LR} \rangle_T).
\end{split}
\end{align}
Note that the second equation follows since $O_{LR}$ takes values in $\{0, 1\}$. This will display a peak, indicating the temperature at which the model is least certain about which phase configurations are from, when $\langle O_{LR} \rangle_T$ crosses $0.5$. $\chi_{kNN}$ is defined similarly. We find evidence that these quantities may also display finite-size scaling behaviour similar to Equations \ref{eqn:susceptibility_fss_secondorder} and \ref{eqn:susceptibility_fss_bkt} which we will use to estimate the critical temperature $T_c$ and the critical exponent of correlation length $\nu$. 

We will initially assume that $\nu$ is known and estimate the critical temperature $T_c$ by fitting the peak temperatures $T_c(L)$ of $\chi_{LR}$ and $\chi_{kNN}$ obtained from multiple lattice sizes to the ansatz
\begin{equation}
\label{eqn:2nd_order_tc_scaling}
    T_c(L) - T_c(\infty) \propto \frac{1}{L^{1/\nu}}.
\end{equation}
in the case of a second order transition, or
\begin{equation}
\label{eqn:bkt_tc_scaling}
    T_c(L) - T_c(\infty) \propto \frac{1}{\log(L)^{1/\nu}}.
\end{equation}
for a BKT transition. 

To estimate $\nu$ (as well as $T_c$), we will use a curve collapse approach, plotting $y = \chi_{LR}$ or $y = \chi_{kNN}$ for multiple lattice sizes simultaneously against $x = L^{1/\nu} \, t $ (second order) or $x = L \, \exp(-bt^{-\nu})$ (BKT) and finding values of $\nu$ and $T_c$ which minimise the distance between the curves using the Nelder-Mead method, as in the procedure described in \cite{Bhattacharjee2001AMO}.

\subsection{Statistical Analysis}
The use of histogram reweighting to extrapolate estimates of ensemble averages to an interval of temperatures around the critical temperature \cite{PhysRevLett.61.2635, PhysRevLett.63.1195} and the use of bootstrap or jackknife analysis to obtain error estimates \cite{efron1979} are standard in quantitative investigations of phase transitions. To provide a full demonstration of a quantitative analysis based on persistent homology we will make use of both techniques which are briefly reviewed in Appendices \ref{appendix:histogram_reweighting} and \ref{appendix:bootstrap}. 

In particular, we use histogram reweighting to interpolate the outputs of our models $\langle O_{LR} \rangle_T$ and $\langle O_{kNN} \rangle_T$. This allows us to obtain interpolated values of the variances $\chi_{LR}$ and $\chi_{kNN}$. Assuming the sampling temperatures are close enough, this allows us to obtain a better estimate of the height and location of peaks of each quantity. 

We estimate the sampling error in the training data and the sampling error in the data in the critical region independently. We do this by performing two bootstrap procedures: the first by resampling the training data, and the second by resampling the data in the critical region. In both cases we resample the data from each temperature individually. The two bootstrap procedures yield approximate sampling distributions of the quantity we are measuring, which we then turn into an error by combining the standard deviations treating the distributions as independent.

\subsection{Analysis Pipeline}
Combining the previous sections we arrive at the procedure for our analysis of each model at each lattice size.
\begin{enumerate}
    \item We sample the model on the given lattice size using the Wolff cluster algorithm \cite{wolff_cluster} at a range of temperatures spanning the phase transition(s). We perform $50,000$ Wolff cluster flips to properly thermalise the model, and $100$ cluster flips between samples to ensure that the autocorrelation is negligible.
    \item For each sample, we compute persistence images with $30 \times 30$ resolution and $\sigma$ equal to $10\%$ of a pixel.
    \item We use persistence images from the low and high temperature phases to train the logistic regression and k-NN models.
    \item Using the trained classification models, we assign a predicted phase to each sample from the critical region.
    \item Close to the peaks in the variances $\chi_{LR}$ and $\chi_{kNN}$ of the classifier we apply multiple histogram reweighting to obtain an interpolated curve and a more precise estimate of the location of the peak.
\end{enumerate}
Once we have the interpolated variance curve and peak temperature for each of the lattice sizes, we estimate $T_c$ and $\nu$ by fitting the peak temperatures to the appropriate finite-size scaling ansatz (Equations \ref{eqn:2nd_order_tc_scaling} and \ref{eqn:bkt_tc_scaling}) and optimising the data collapse of the variance curves. For each lattice size we perform two bootstraps: first by resampling the training samples, and second by resampling the samples in the critical region. In each case we resample $500$ times, obtaining bootstrap distributions for the estimates of $T_c$ and $\nu$. We estimate the error in these quantities by taking the square root of the sum of the variances of the two bootstrap distributions.

\section{Analysis}
\label{sec:analysis}

We analyse three different variants of the 2-dimensional XY model; each undergoes a Berezinskii-Kosterlitz-Thouless (BKT) phase transition. One of the variants also exhibits a second order transition in the Ising universality class, and it  presents an interesting challenge to classify both transitions. For each model, we considered square lattices with periodic boundary conditions and linear sizes $L = 30$, $40$, $50$, $60$, $70$, $80$, $100$, $120$, $140$.

\subsection{XY Model}
The $2$-dimensional XY model is defined on an $L\times L$ square lattice $\Lambda$ by assigning an angle $\theta_i \in S^1$ to each lattice site $i \in \Lambda$. The energy of a given configuration of spins $\boldsymbol\theta = \{\theta_i\}_{i \in \Lambda}$ is given by the Hamiltonian $$H(\boldsymbol\theta) = - J \sum_{\langle ij \rangle}\cos(\theta_i - \theta_j)$$ where $\langle ij \rangle$ ranges over neighbouring lattice sites and $J$ is a coupling parameter we shall set equal to $1$. At low temperatures spins tend to align with their neighbours, but collectively twist in spin waves preventing true long-range order. Moreover a small number of vortices and antivortices, where spins twist round the full circle, may be found in bound pairs. As the temperature increases, the model undergoes a BKT transition driven by the unbinding of these vortex-antivortex pairs, so that at high temperatures lone (anti)vortices proliferate. The critical temperature is approximately $T = 0.8929$ \cite{Hasenbusch2005TheTX} and the critical exponent of correlation length is $\nu = \frac{1}{2}$.

\begin{figure}[h]
    \centering
    \scalebox{0.34}{\includegraphics{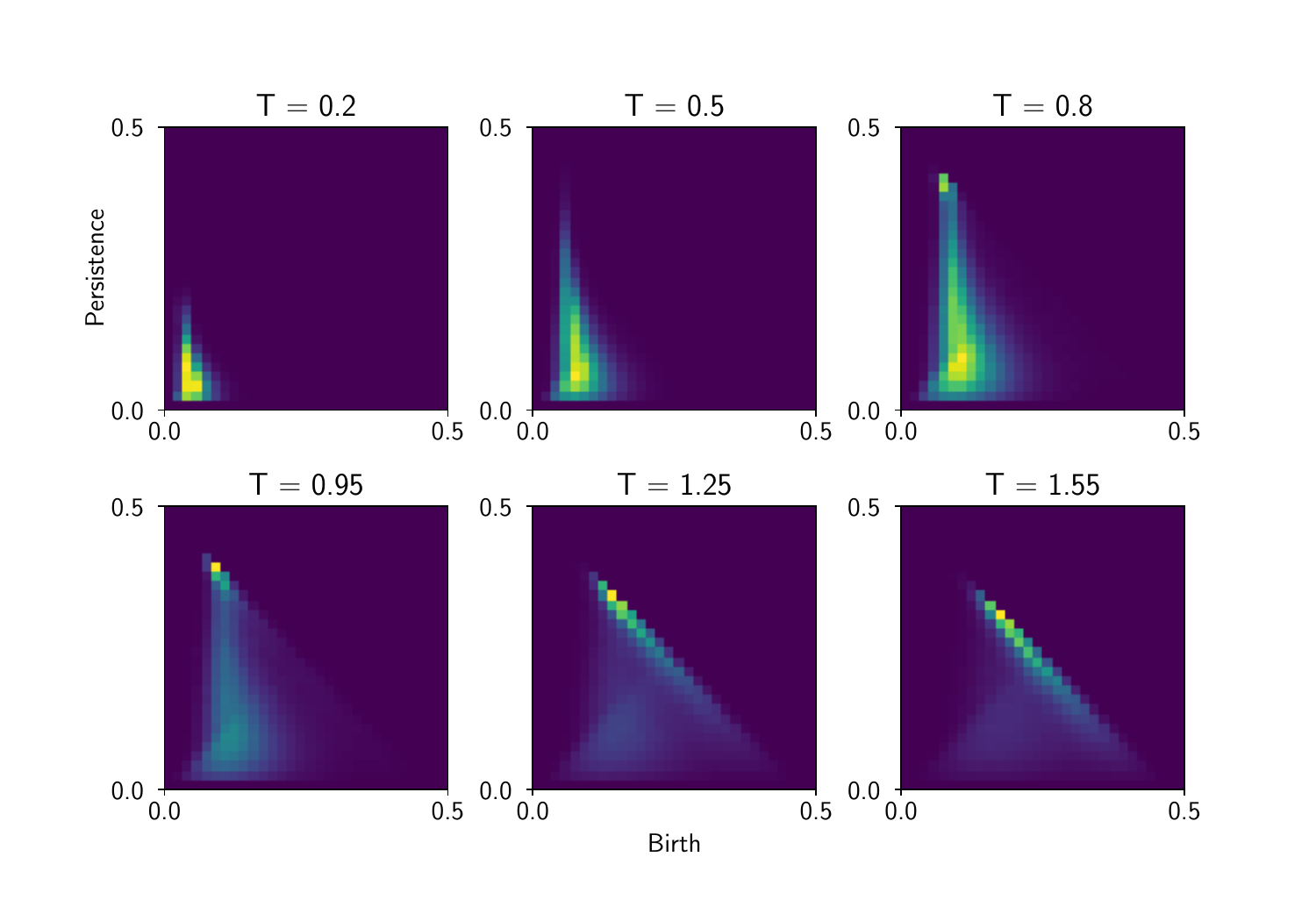}}
    \caption{The average $H_1$ persistence images in birth-persistence coordinates at different temperatures for the XY model with $L = 30$.}
    \label{fig:xy_mypis}
\end{figure}

\begin{figure*}[t]
    \centering
    \scalebox{0.45}{\includegraphics{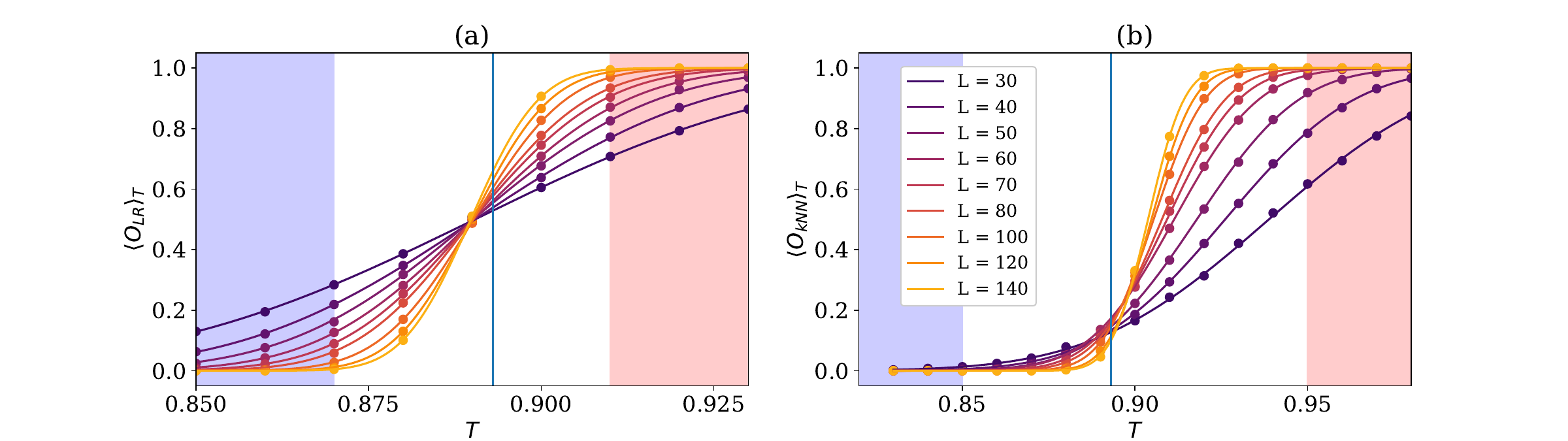}}
    \caption{Plots showing (a) $\langle O_{LR} \rangle$ and (b) $\langle O_{kNN} \rangle$ as a function of temperature for each lattice size for the XY model. The shaded regions indicate the temperatures used for the low and high temperature training data. The vertical line shows the location of the expected critical temperature $T_c = 0.8929$.}
    \label{fig:xy_curves}
\end{figure*}

\begin{figure*}[t]
    \centering
    \scalebox{0.36}{\includegraphics{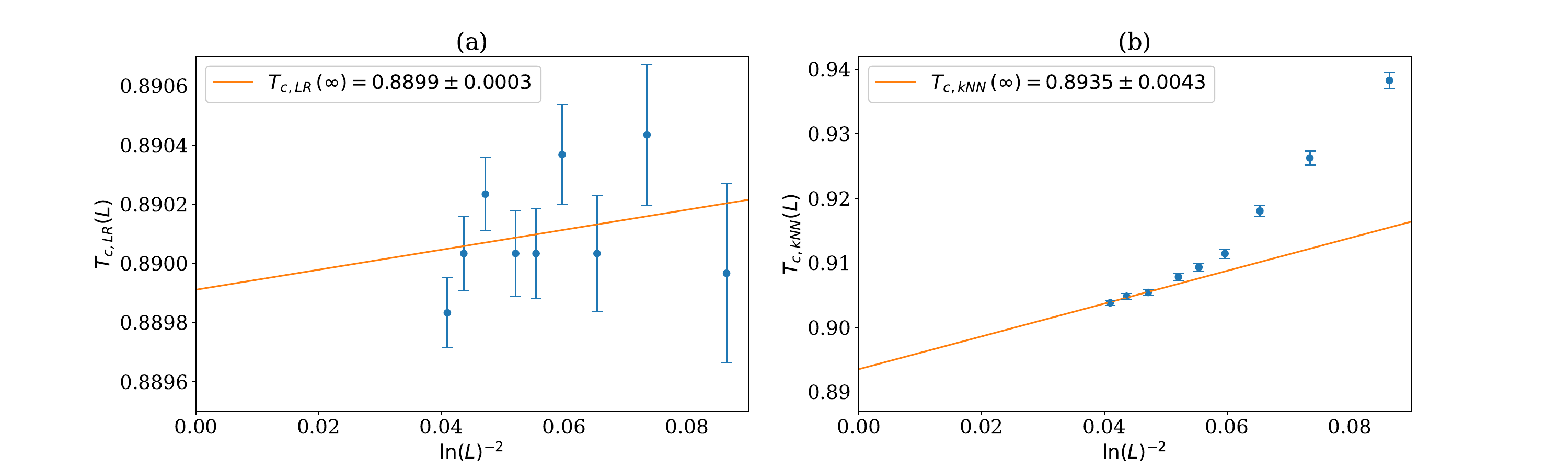}}
    \caption{Estimating the critical temperature for the XY model using (a) logistic regression and (b) k-nearest neighbours. The pseudo-critical temperatures for the different lattice sizes, calculated from finding the peak of $\chi_p$, are fitted to the ansatz in Equation \ref{eqn:bkt_tc_scaling}. For the logistic regression we use all the lattice sizes in the fit, and for the k-nearest neighbours we use the largest three lattice sizes. The intercept gives the estimate for $T_c(\infty)$. Error bars are estimated by bootstrapping.}
    \label{fig:xy_Tc}
\end{figure*}

Using the angle difference filtration described in Section \ref{sec:filtration} we obtain average persistence images as shown in Figure \ref{fig:xy_mypis}. At low temperatures we see that most points in the persistence diagrams are concentrated in the lower left corner. These come from the presence of spin waves: spins tend to differ more with those in the opposite corner of a plaquette than with their immediate neighbours, producing a short-lived cycle. As the temperature increases we observe that the spin-wave cycles persist longer and longer. At around $T = 0.8$, $0.95$, close to the critical point, we begin to see points close to the downwards diagonal $\mathit{persistence} = 0.5 - \mathit{birth}$, or equivalently $\mathit{death} = 0.5$. These represent (anti)vortices: they are born reasonably early, as spins far away from the centre vary slowly, but die much later due to the large difference in spins at the vortex core. In fact, we can check that the sum of the components of the persistence image lying on the diagonal and the two immediate subdiagonals correlates well with the absolute vorticity (the total count of vortices and antivortices) of the configurations. For example, computing the Pearson correlation coefficient on $2000$ configurations at $T = 1.0$ for $L = 140$ yields a correlation coefficient of $r = 0.70$, $p < 0.001$. At high temperatures we see this concentration of cycles on the diagonal increase and shift rightwards, indicating a disordered phase with many vortices.
\begin{figure}[h]
    \centering
    \scalebox{0.47}{\includegraphics{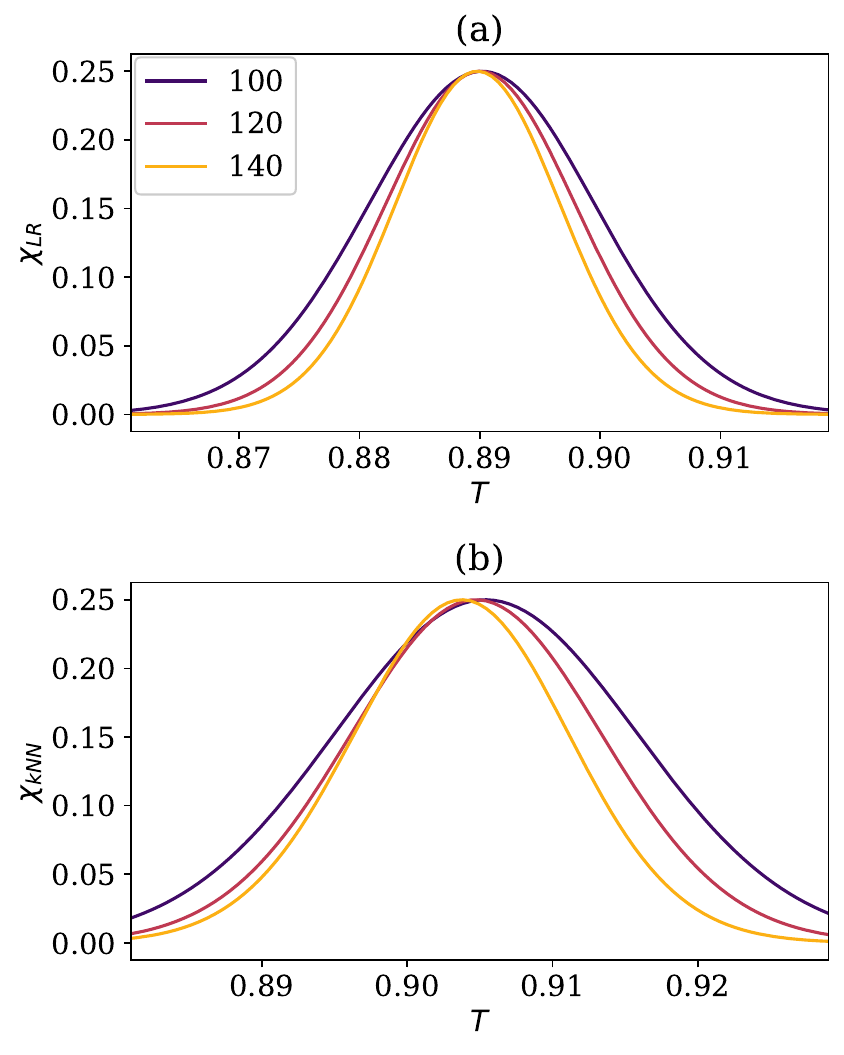}}
    \caption{Plots showing (a) $\chi_{LR}$ and (b) $\chi_{kNN}$ as a function of temperature for the largest three lattice sizes of the XY model. These are what we use to perform the curve collapse procedure.}
    \label{fig:xy_var_curves}
\end{figure}
\subsubsection{Logistic Regression Analysis}
\begin{figure}[h]
    \centering
    \scalebox{0.44}{\includegraphics{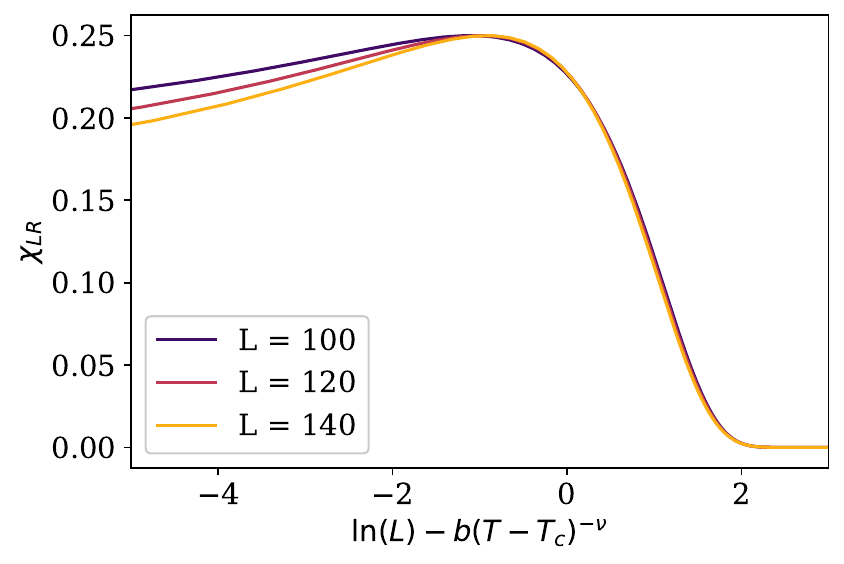}}
    \caption{The curve collapse of $\chi_{LR}$ for the XY model with $T_c = 0.8824$, $\nu = 0.4968$ and $b = 0.5098$.}
    \label{fig:xy_lr_collapse}
\end{figure}

We trained logistic regression models on samples drawn from $T = 0.85$, $0.86$ and $0.87$ in the low temperature phase, and $T = 0.91$, $0.92$ and $0.93$ in the high temperature phase with $10,000$ samples from each. The regularisation hyper-parameter was set to $C = 0.001$. We evaluated the models with $10,000$ samples from each of $T = 0.88$, $0.89$ and $0.90$. A plot of the resulting phase indicators is shown in Figure \ref{fig:xy_curves} and their variance curves are shown in Figure \ref{fig:xy_var_curves}. The plot of the pseudo-critical temperatures against $log(L)^{-2}$ is shown in Figure \ref{fig:xy_Tc}. We do not observe any significant lattice-size dependence in the pseudo-critical temperatures. They instead seem to be distributed close to $T = 0.89$ which is the midpoint of the training temperatures. A straight line fit yields an extrapolated critical temperature of $$T_c = 0.8872 \pm 0.0009,$$ well below the expected $T_c = 0.8929$. The curve collapse (Figure \ref{fig:xy_lr_collapse}) procedure gives
\begin{center}
\begin{tabular}{ c }
 $T_c = 0.8824 \pm 0.0001$ \\ $\nu = 0.4968 \pm 0.0055$ \\ $b = 0.5098 \pm 0.0068,$
\end{tabular}
\end{center}
not accounting within one standard deviation for the expected values of $T_c = 0.8929$ and $\nu = \frac{1}{2}$. 

\begin{figure}[h]
    \centering
    \scalebox{0.5}{\includegraphics{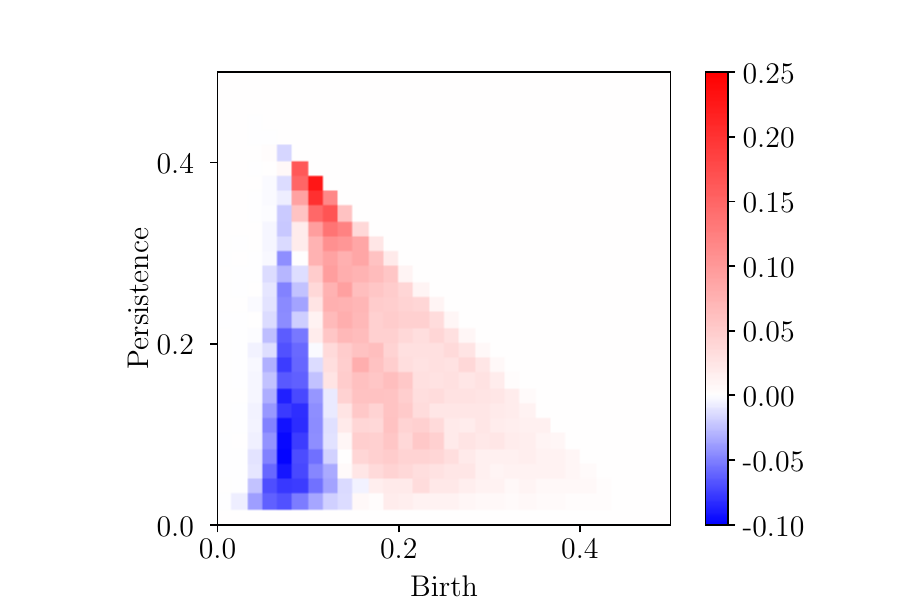}}
    \caption{The weights of the logistic regression model trained on the XY model configurations with $L = 140$.}
    \label{fig:xy_coefs}
\end{figure}

An advantage of using a generalised linear model like logistic regression, as explored in \cite{cole2020quantitative}, is that we can easily match the learned weights against the pixels of the persistence images. This allows us to interpret how the classifier distinguishes phases. The weights of the logistic regression model trained on the $L = 140$ XY model data is shown in Figure \ref{fig:xy_coefs}. We see that the low temperature phase is characterised by cycles which are born early and which tend to have low persistence, representing spin waves. The high temperature phase is indicated by cycles with a later birth time and persistence. In particular, the most important region in identifying the high temperature phase is close to $\mathit{birth} = 0.1$, $\mathit{persistence} = 0.4$ which detects (anti)vortex cycles beginning to change behaviour and move down the diagonal $\mathit{persistence} = 0.5 - \mathit{birth}$.

\begin{figure}[h]
    \centering
    \scalebox{0.44}{\includegraphics{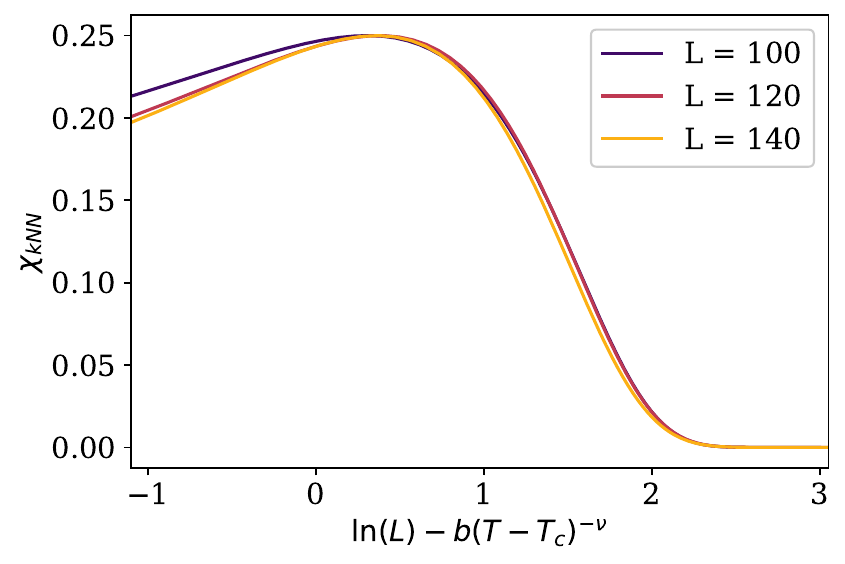}}
    \caption{The curve collapse of $\chi_{kNN}$ for the XY model with $T_c = 0.8918$, $\nu = 0.4972$ and $b = 0.5073$.}
    \label{fig:xy_knn_collapse}
\end{figure}

\subsubsection{k-Nearest Neighbours Analysis}

In the case of the XY model, we found that the k-nearest neighbours classification worked best when trained on a broad range of temperatures. We trained the models on samples drawn from $T = 0.20$, $0.25$, \ldots, $0.85$ in the low temperature phase, and $T = 0.95$, $1.00$, \ldots, $1.60$ in the high temperature phase with $2000$ samples from each. The neighbours hyper-parameter was set to $k = 30$. We evaluated the models with $10,000$ samples from each of $T = 0.90$, $0.905$, \ldots, $0.95$. A plot of the resulting phase indicators is shown in Figure \ref{fig:xy_curves}. The plot of the pseudo-critical temperatures against $\log(L)^{-2}$ is shown in Figure \ref{fig:xy_Tc}. Here we see an asymptotic convergence towards a linear dependence between the pseudo-critical temperatures $T_c(L)$ and $\log(L)^{-2}$. Fitting a straight line to the largest three lattice sizes yields $$T_c = 0.8935 \pm 0.0043,$$ much closer to the expected $T_c \approx 0.8929$ than the result of the logistic regression approach. The curve collapse (Figure \ref{fig:xy_knn_collapse}) procedure gives
\begin{center}
\begin{tabular}{ c }
 $T_c = 0.8918 \pm 0.0033$ \\ $\nu = 0.4972 \pm 0.0264$ \\ $b = 0.5073 \pm 0.0137,$
\end{tabular}
\end{center}
very close to the expected values.

\subsection{Constrained XY Model}

What we will refer to as the $2$-dimensional Constrained XY model was introduced and investigated in \cite{Bietenholz2010TopologicalLA, Bietenholz2013TopologicalLA} where it is called an XY model with a topological lattice action. It is defined similarly to the classical XY model by assigning an angle $\theta_i \in S^1$ to each lattice site $i \in \Lambda$ of an $L \times L$ square lattice $\Lambda$. However the Hamiltonian is defined as
$$H(\boldsymbol\theta) =
     \begin{cases}
       0 &\quad\text{if }\frac{1}{2\pi}{\vert\theta_i-\theta_j\vert} \leq \delta \text{ for all }\langle i, j \rangle\\
       \infty &\quad\text{otherwise.}\\
     \end{cases}$$
Therefore all configurations are constrained so that the spins at neighbouring sites cannot differ by more than $\delta$. Since the partition function does not depend on the thermodynamic temperature, we consider the parameter $\delta$ as taking on this role instead and the model undergoes a BKT transition as $\delta$ increases at approximately $\delta = 0.2825$ \cite{Bietenholz2013TopologicalLA} with $\nu = \frac{1}{2}$. Notice that while $\delta < 0.25$ no (anti)vortices may form. 

\begin{figure}[h]
    \centering
    \scalebox{0.34}{\includegraphics{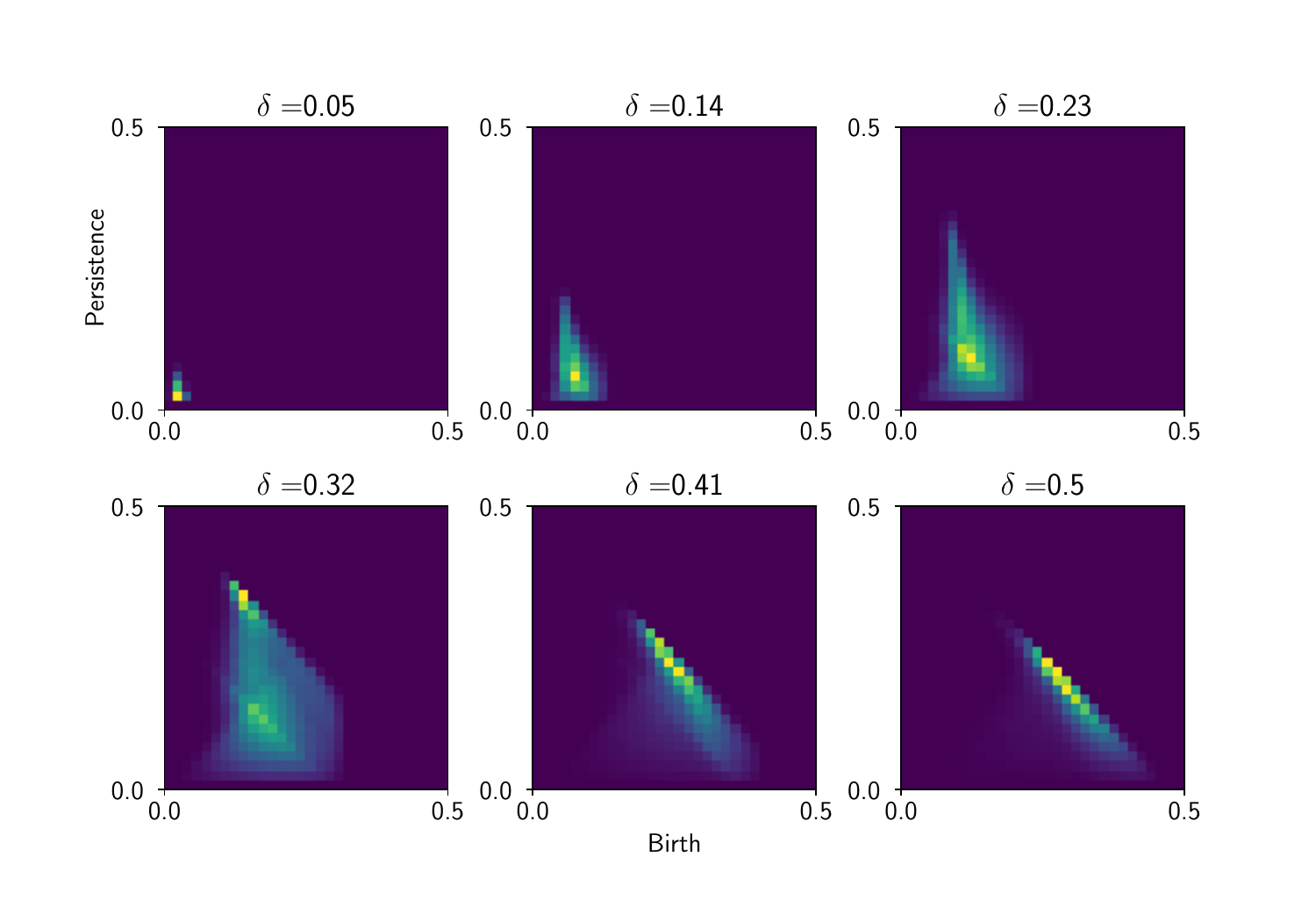}}
    \caption{The average $H_1$ persistence image in birth-persistence coordinates at different deltas for the Constrained XY model with $L = 30$.}
    \label{fig:cxy_mypis}
\end{figure}

Using the angle difference filtration described in Section \ref{sec:filtration} we obtain average persistence images as shown in Figure \ref{fig:cxy_mypis}. We immediately see a resemblance with the persistence images obtained for the XY model in Figure \ref{fig:xy_mypis} except that we see a cutoff effect at $\mathit{birth} = \delta$, since by this point all neighbouring lattice sites must have been connected in the filtration. For this model we must adjust our methodology slightly since histogram reweighting is not possible. Instead we will sample deltas more densely, then to extract the maximums of $\chi_{LR}$ and $\chi_{kNN}$ we will fit a parabola to the three highest points.

\begin{figure*}[t]
    \centering
    \scalebox{0.45}{\includegraphics{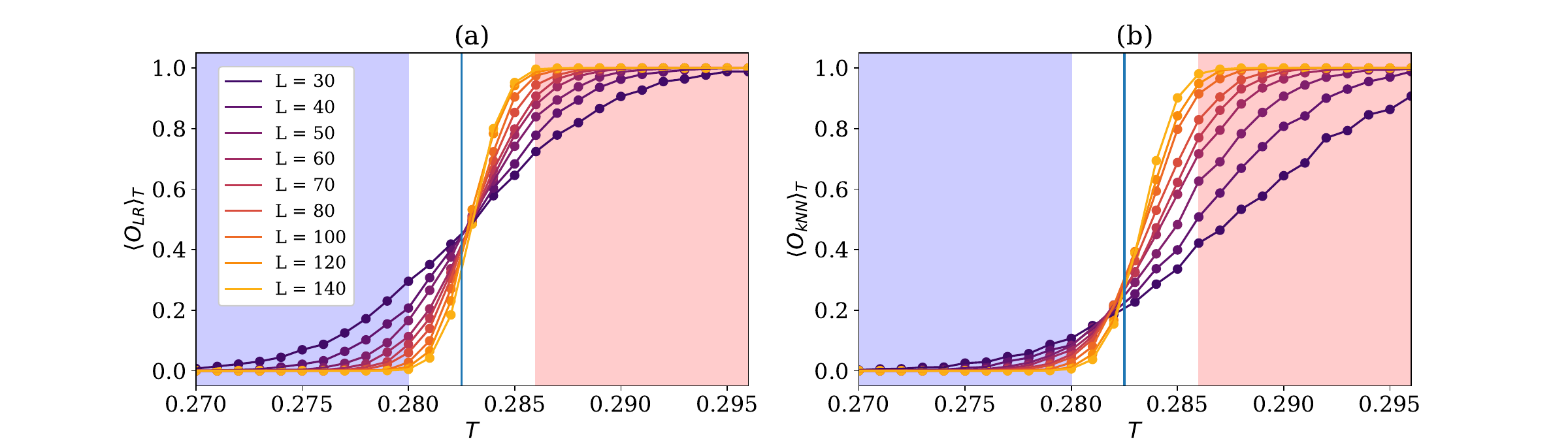}}
    \caption{Plots showing (a) $\langle O_{LR} \rangle$ and (b) $\langle O_{kNN} \rangle$ as a function of delta for each lattice size for the Constrained XY model. The shaded regions indicate the deltas used for the low and high delta training data. The vertical line shows the location of the expected critical delta $\delta_c = 0.2825$. Note that for the k-NN plot the training regions extend further away than what is shown.}
    \label{fig:cxy_curves}
\end{figure*}

\begin{figure*}[t]
    \centering
    \scalebox{0.36}{\includegraphics{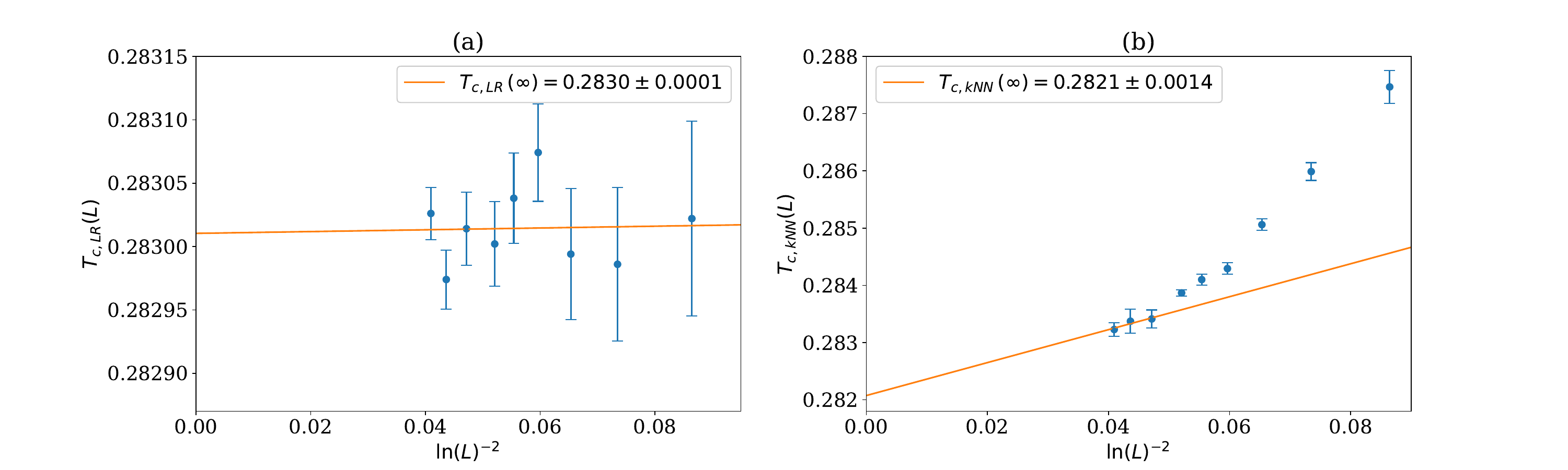}}
    \caption{Estimating the critical delta for the Constrained XY model using (a) logistic regression and (b) k-nearest neighbours. The pseudo-critical deltas for the different lattice sizes, calculated from finding the peak of $\chi_p$, are fitted to the ansatz in Equation \ref{eqn:bkt_tc_scaling}. For the logistic regression we use all the lattice sizes in the fit, and for the k-nearest neighbours we use the largest three lattice sizes. The intercept gives the estimate for $\delta_c(\infty)$. Error bars are estimated by bootstrapping.}
    \label{fig:cxy_Tc}
\end{figure*}

\subsubsection{Logistic Regression Analysis}

\begin{figure*}[h!t]
    \centering
    \begin{minipage}[b]{.4\textwidth}
    \scalebox{0.44}{\includegraphics{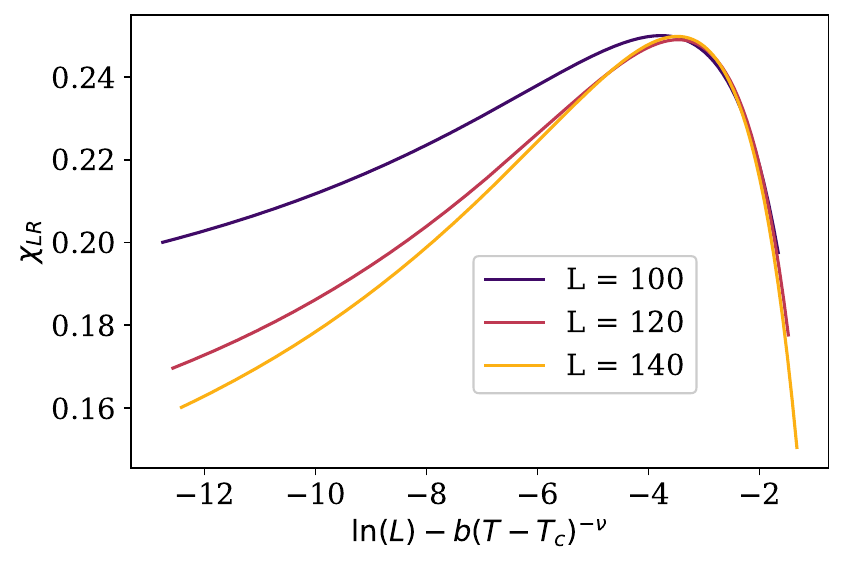}}
    \caption{The curve collapse of $\chi_{LR}$ for the Constrained XY model with $\delta_c = 0.2843$, $\nu = 0.4999$ and $b = 0.3009$.}
    \label{fig:cxy_lr_collapse}
    \end{minipage}\qquad\quad
    \begin{minipage}[b]{.4\textwidth}
    \scalebox{0.5}{\includegraphics{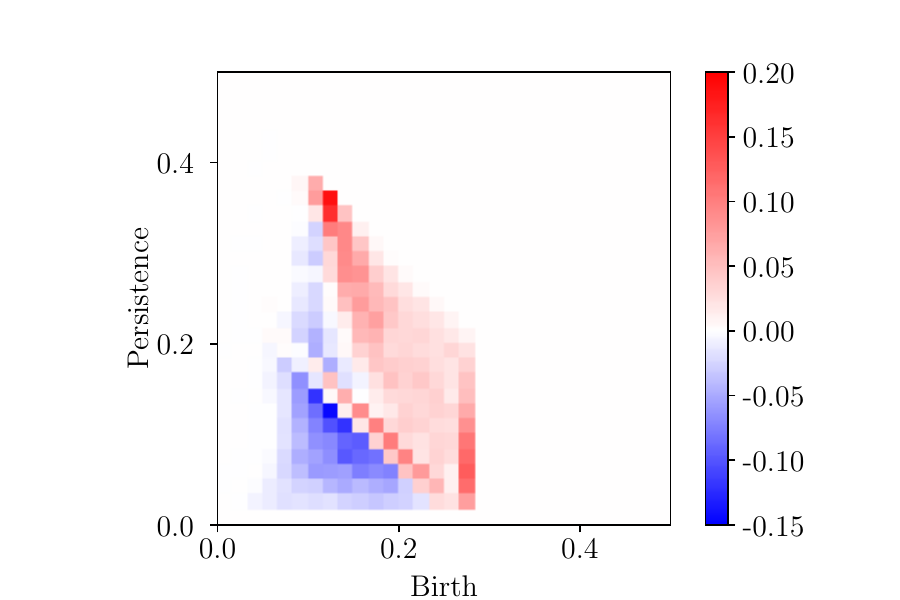}}
    \caption{The weights of the logistic regression model trained on configurations over the BKT transition in the Constrained XY model with $L = 140$.}
    \label{fig:cxy_coefs}
    \end{minipage}
\end{figure*}

We trained logistic regression models on samples drawn from $\delta = 0.27$, $0.272$, \ldots, $0.28$ in the low delta phase, and $\delta = 0.286$, $0.288$, \ldots, $0.296$ in the high delta phase with $4000$ samples from each. The regularisation hyper-parameter was set to $C = 0.001$. We evaluated the models with $4000$ samples from each of $\delta = 0.27$, $0.271$, \ldots, $0.296$. A plot of the resulting phase indicators is shown in Figure \ref{fig:cxy_curves}. The plot of the pseudo-critical deltas against $\log(L)^{-2}$ is shown in Figure \ref{fig:cxy_Tc}. We do not observe any significant lattice-size dependence in the pseudo-critical deltas. They instead seem to be distributed close to $\delta = 0.283$ which is the midpoint of the training deltas. The curve collapse (Figure \ref{fig:cxy_lr_collapse}) procedure gives
\begin{center}
\begin{tabular}{ c }
 $\delta_c = 0.2843 \pm 0.0013$ \\ $\nu = 0.4999 \pm 0.0189$ \\ $b = 0.3009 \pm 0.0041,$
\end{tabular}
\end{center}
which is not likely to account for the expected value of $\delta_c = 0.2825$ but does support $\nu = \frac{1}{2}$. 

The weights of the logistic regression model trained for $L = 140$ are shown in Figure \ref{fig:cxy_coefs}. We observe a similarity to the weights learnt for the XY model in Figure \ref{fig:xy_coefs} although in this case it appears to be more difficult to delineate which regions of the persistence images indicate the two phases.

\subsubsection{k-Nearest Neighbours Analysis}

We trained the k-nearest neighbours models on samples drawn from $\delta = 0.27$, $0.272$, \ldots, $0.28$ in the low delta phase, and $\delta = 0.286$, $0.288$, \ldots, $0.296$ in the high delta phase with $4000$ samples from each. The neighbours hyper-parameter was set to $k = 30$. We evaluated the models with $4000$ samples from each of $\delta = 0.27$, $0.271$, \ldots, $0.296$. A plot of the resulting phase indicators is shown in Figure \ref{fig:cxy_curves}. The plot of the pseudo-critical deltas against $\log(L)^{-2}$ is shown in Figure \ref{fig:cxy_Tc}. Here we see an asymptotic convergence towards a linear dependence between the pseudo-critical deltas $\delta_c(L)$ and $\log(L)^{-2}$. Fitting a straight line to the largest three lattice sizes yields $$\delta_c = 0.2821 \pm 0.0014.$$

\begin{figure}[h]
    \centering
    \scalebox{0.44}{\includegraphics{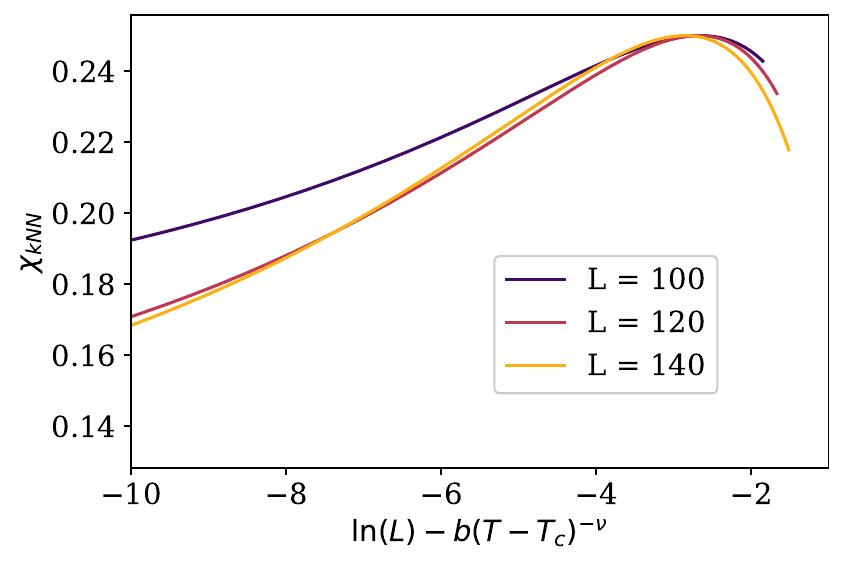}}
    \caption{The curve collapse of $\chi_{kNN}$ for the Constrained XY model with $\delta_c = 0.2818$, $\nu = 0.5003$ and $b = 0.3022$.}
    \label{fig:cxy_knn_collapse}
\end{figure}

The curve collapse (Figure \ref{fig:cxy_knn_collapse}) procedure gives
\begin{center}
\begin{tabular}{ c }
 $\delta_c = 0.2818 \pm 0.0017$ \\ $\nu = 0.5003 \pm 0.0206$ \\ $b = 0.5022 \pm 0.0048,$
\end{tabular}
\end{center}
very close to the expected values.

\subsection{Nematic XY Model}
\label{sec:nematic}
\begin{figure}[h]
    \centering
    \scalebox{0.6}{\includegraphics{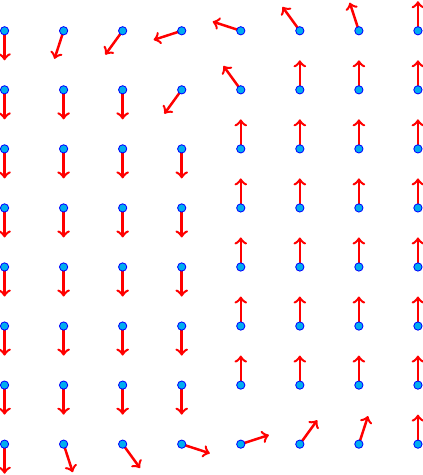}}
    \caption{A configuration with a vortex that has stretched out into two half-vortices separated by a domain wall.}
    \label{fig:domain}
\end{figure}

There are a variety of generalised XY models with nematic interactions. We will consider the model with Hamiltonian $$H(\boldsymbol\theta) = - \sum_{\langle ij \rangle} \big[ \Delta\, cos(\theta_i - \theta_j) + (1 - \Delta)\, cos(2(\theta_i - \theta_j)) \big]$$ where we will fix $\Delta = 0.15$. The first term is the usual XY interaction, but the second term is a nematic interaction which remains invariant when any individual spin is rotated 180 degrees. We can imagine this as an interaction between the spins considered as headless rods: spins which are parallel contribute less energy, even if they point in opposite directions. The $T$-$\Delta$ phase diagram of this model is explored in \cite{Nui2018CorrelationLI, Serna:2017ura, PhysRevE.94.032140}, and we see that at our chosen $\Delta = 0.15$, it undergoes two phase transitions as temperature increases. The first is an Ising-type transition from a magnetic phase to a nematic phase at $T \approx 0.3314$ (as estimated using the magnetic susceptibility) resulting in (anti)vortices (which remain bound into vortex-antivortex pairs) stretching into domain walls with a half-(anti)vortex at each end; across the wall the spins flip by $\pi$. See Figure \ref{fig:domain} for an example. The second is a BKT transition to a paramagnetic phase at $T \approx 0.7808$ (as estimated using the magnetic susceptibility) driven by the unbinding of these pairs of now-elongated vortices and antivortices.

\begin{figure}[h!]
    \centering
    \scalebox{0.34}{\includegraphics{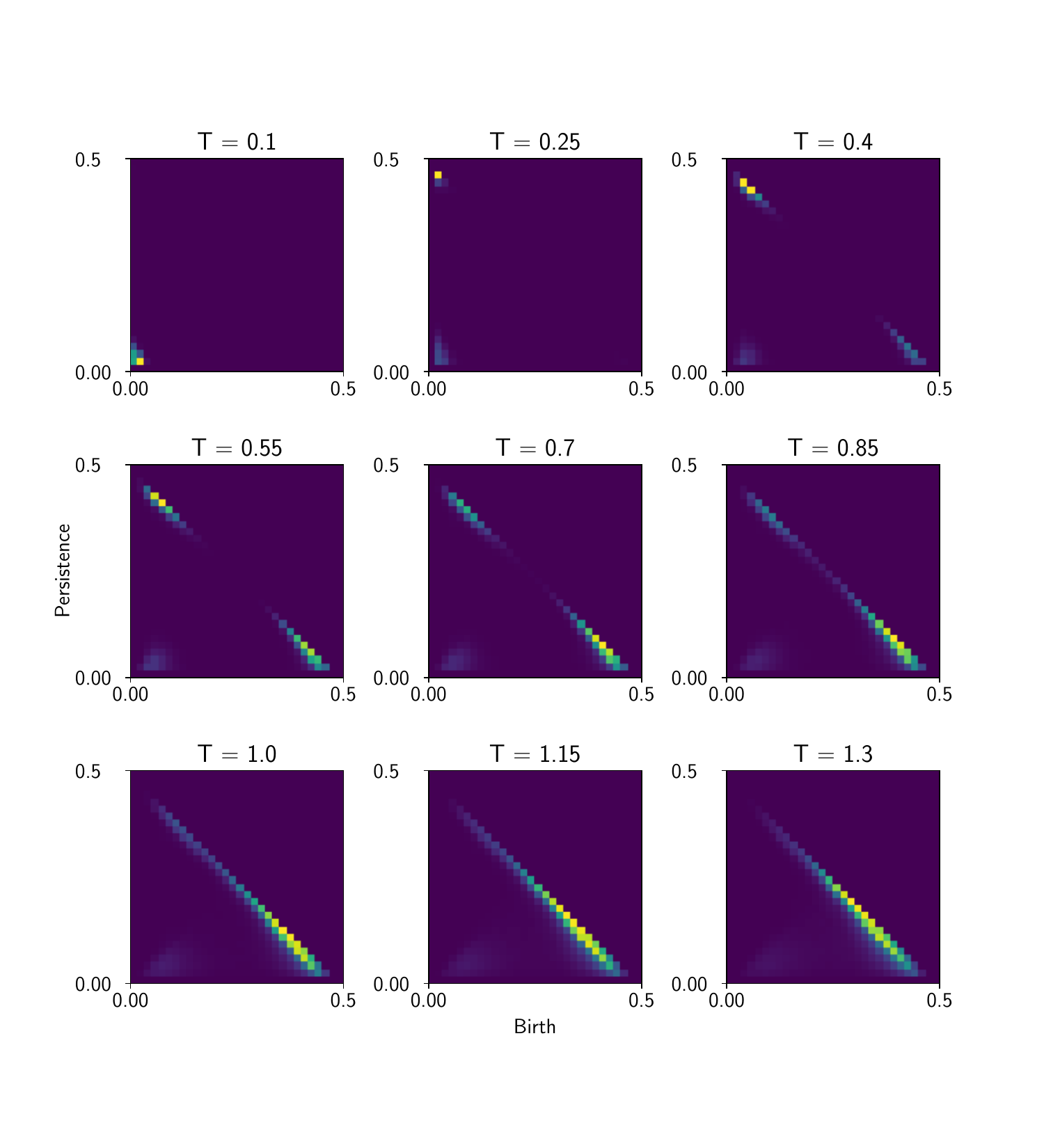}}
    \caption{The average $H_1$ persistence image in birth-persistence coordinates at different temperatures for the Nematic XY model with $L = 30$ using the angle difference filtration. The Magnetic-Nematic phase transition occurs between the middle and right images on the top row and the Nematic-Paramagnetic BKT transition occurs between the middle and right images on the middle row.}
    \label{fig:nxy_mypis}
\end{figure}

\begin{figure}[h!]
    \centering
    \scalebox{0.34}{\includegraphics{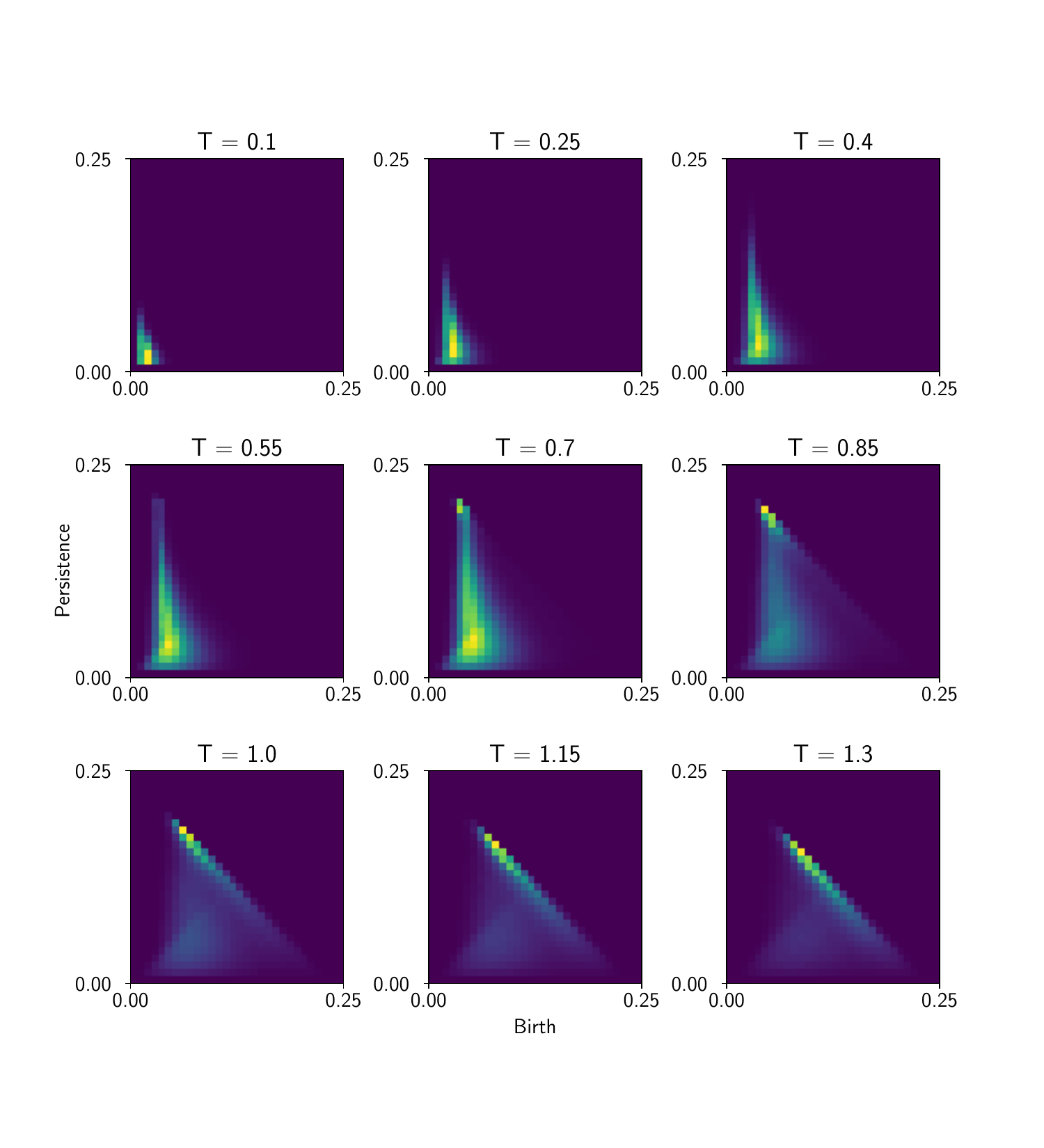}}
    \caption{The average $H_1$ persistence image in birth-persistence coordinates at different temperatures for the Nematic XY model with $L = 30$ using the nematic angle difference filtration. The Magnetic-Nematic phase transition occurs between the middle and right images on the top row and the Nematic-Paramagnetic BKT transition occurs between the middle and right images on the middle row. Note the similarity with Figures \ref{fig:xy_mypis} and \ref{fig:cxy_mypis}.}
    \label{fig:nxy_mypisn}
\end{figure}

Following the intuition developed in Section \ref{sec:filtration}, we consider two different filtrations: The first is the angle difference filtration used for the XY and Constrained XY model, where each edge $\langle i j \rangle$ of the lattice is added into the filtration at time $\frac{1}{2\pi}{d_{ij}}$; The second is the nematic angle difference filtration which uses the nematic distance between spins, adding in edges at time $\frac{1}{2\pi}d_{ij}^n = \min(\frac{1}{2\pi}d_{ij}, 0.5 - \frac{1}{2\pi}d_{ij})$. The resulting average persistence images are shown in Figures \ref{fig:nxy_mypis} and \ref{fig:nxy_mypisn} respectively. 

From Figure \ref{fig:nxy_mypis} we see that the Magnetic-Nematic transition is manifested in the angle difference filtration by the emergence of a cluster in the bottom right of the persistence image and the rightwards movement of the cluster in the top left. These correspond to the appearance of domain walls in configurations. In particular, at a time close to $0.5$ in the filtration, the edges which cross domain walls will get added all at once, forming many short-lived cycles. Meanwhile, (anti)vortices get stretched out into strings so that more spins must be connected in the filtration before a hole is formed, generally causing the time at which this happens to increase a little. There is little qualitative difference between the images across the BKT transition however. In Figure \ref{fig:nxy_mypisn} we see a familiar picture of the BKT transition which is very similar to that observed in the XY model and Constrained XY model, while the Ising-type transition is not detectable at all. We also looked at a combined angle difference filtration using $\frac{\Delta}{2\pi} d_{ij} + \frac{1 - \Delta}{2\pi} d_{ij}^n$, but while this did seem to detect both phase transitions, it was difficult to effectively train the classification models to identify two phases at a time.

\begin{figure*}[t]
    \centering
    \scalebox{0.45}{\includegraphics{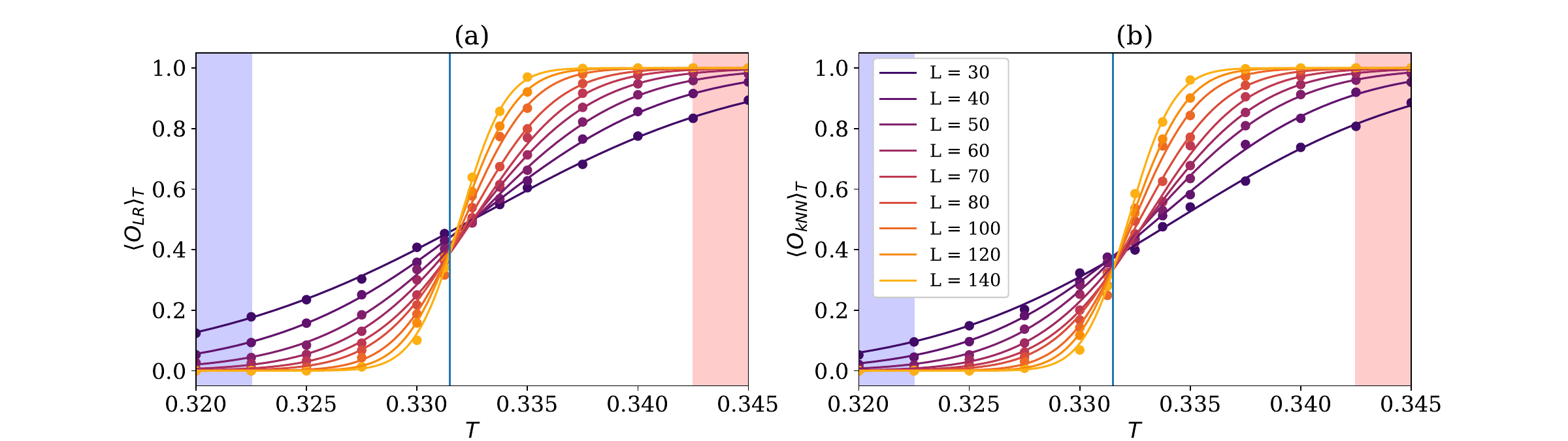}}
    \caption{Plots showing (a) $\langle O_{LR} \rangle$ and (b) $\langle O_{kNN} \rangle$ as a function of temperature for each lattice size for the Magnetic-Nematic transition in the Nematic XY model. The shaded regions indicate the temperatures used for the low and high temperature training data. The vertical line shows the location of the expected critical temperature $T_c = 0.3314$.}
    \label{fig:nxy_ising_curves}
\end{figure*}

\subsubsection{Logistic Regression Analysis of Magnetic-Nematic Transition}

\begin{figure*}[t]
    \centering
    \scalebox{0.36}{\includegraphics{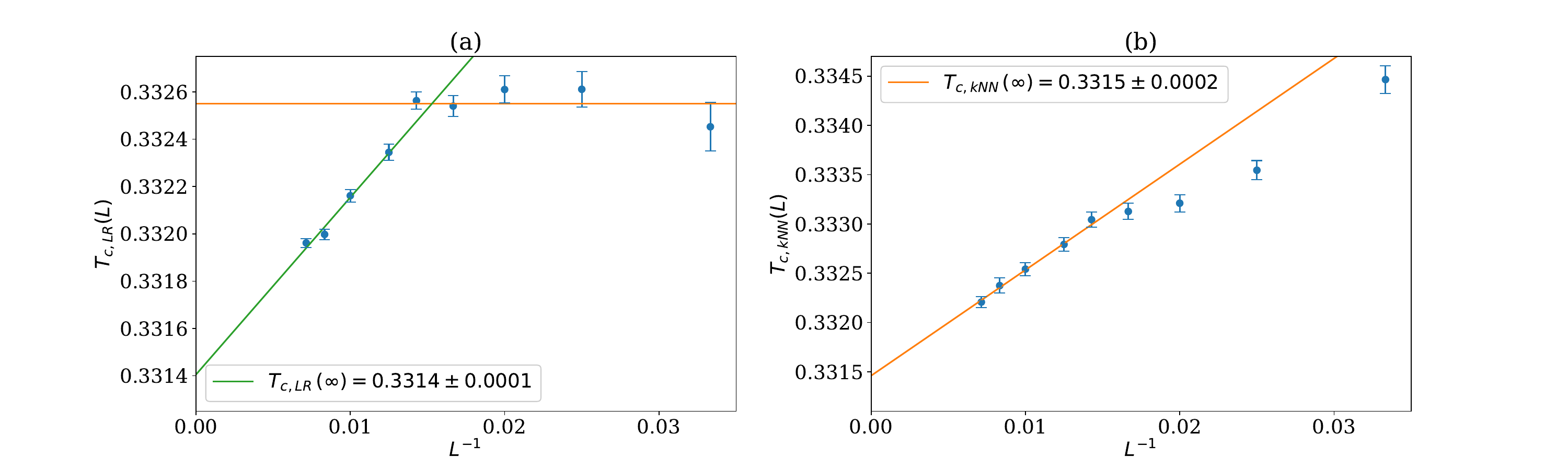}}
    \caption{Estimating the critical temperature for the Magnetic-Nematic transition in the Nematic XY model using (a) logistic regression and (b) k-nearest neighbours. The pseudo-critical temperatures for the different lattice sizes, calculated from finding the peak of $\chi_p$, are fitted to the ansatz in Equation \ref{eqn:2nd_order_tc_scaling}. In both cases we use the largest four lattice sizes for the fit. The intercept gives the estimate for $T_c(\infty)$. Error bars are estimated by bootstrapping.}
    \label{fig:nxy_ising_Tc}
\end{figure*}

\begin{figure*}[h!t]
    \centering
    \begin{minipage}[b]{.4\textwidth}
    \scalebox{0.44}{\includegraphics{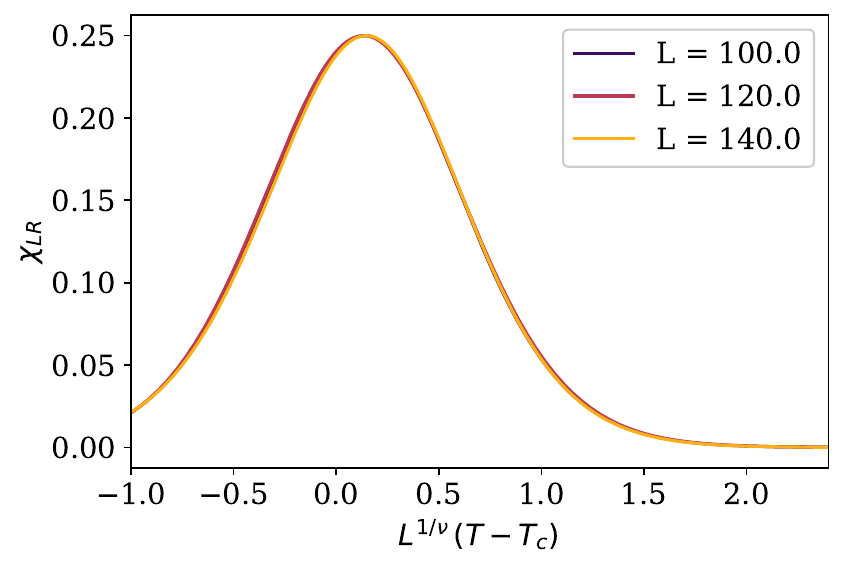}}
    \caption{The curve collapse of $\chi_{LR}$ for the Magnetic-Nematic transition in the Nematic XY model with $T_c = 0.3315$ and $\nu = 0.8562$.}
    \label{fig:nxy_ising_lr_collapse}
    \end{minipage}\qquad\quad
    \begin{minipage}[b]{.4\textwidth}
    \scalebox{0.5}{\includegraphics{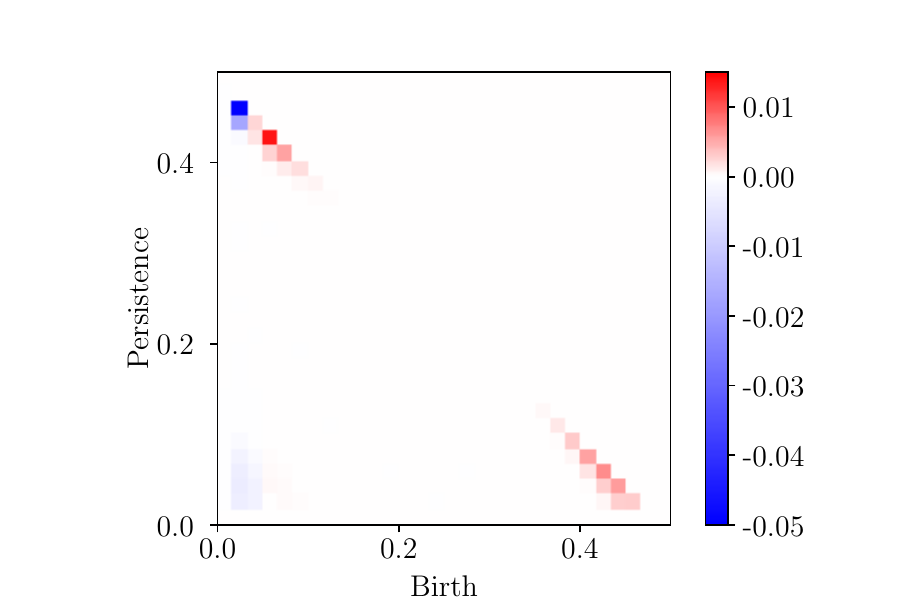}}
    \caption{The weights of the logistic regression model trained on configurations over the Magnetic-Nematic transition in the Nematic XY model with $L = 140$.}
    \label{fig:nxy_ising_coefs}
    \end{minipage}
\end{figure*}

We trained logistic regression models on samples drawn from $T = 0.32$ and $0.3225$ in the low temperature phase, and $T = 0.3425$ and $0.345$ in the high temperature phase with $10,000$ samples from each. The regularisation hyper-parameter was set to $C = 10^{-6}$. We evaluated the models with $10,000$ samples from each of $T = 0.33$, $0.33125$, \ldots, $0.335$. A plot of the resulting phase indicators is shown in Figure \ref{fig:nxy_ising_curves}. The plot of the pseudo-critical temperatures against $L^{-1}$ is shown in Figure \ref{fig:nxy_ising_Tc}. For the lower lattice sizes $L < 60$, we do not observe any significant lattice-size dependence in the pseudo-critical temperatures. They instead seem to be distributed close to $T = 0.3325$ which is the midpoint of the training temperatures. At the larger lattice sizes $L \geq 60$, a linear dependence on $L^{-1}$ emerges. Fitting a line to the largest four lattice sizes yields an extrapolated critical temperature of $$T_c = 0.3314 \pm 0.0001.$$ The curve collapse (Figure \ref{fig:nxy_ising_lr_collapse}) procedure gives
\begin{center}
\begin{tabular}{ c }
 $T_c = 0.3315 \pm 0.0001$ \\ $\nu = 0.8562 \pm 0.0102.$
\end{tabular}
\end{center}

While these estimates of the critical temperature are good, the expected value of $\nu = 1$ doesn't fall within the error bars estimated with this approach. 

The weights of the logistic regression model trained for $L = 140$ are shown in Figure \ref{fig:nxy_ising_coefs}. We observe that the classifier learns to detect exactly what we saw in Figure \ref{fig:nxy_mypis}, namely a rightwards shift of the upper left cluster, and the emergence of a cluster in the bottom right, corresponding to domain walls forming in the configurations.

\subsubsection{k-Nearest Neighbours Analysis of Magnetic-Nematic Transition}

We trained the k-nearest neighbours models on samples drawn from $T = 0.32$ and $0.3225$ in the low temperature phase, and $T = 0.3425$ and $0.345$ in the high temperature phase with $4000$ samples from each. The neighbours hyper-parameter was set to $k = 30$. We evaluated the models with $4000$ samples from each of $T = 0.33$, $0.33125$, \ldots, $0.335$. A plot of the resulting phase indicators is shown in Figure \ref{fig:nxy_ising_curves}. The plot of the pseudo-critical temperatures against $L^{-1}$ is shown in Figure \ref{fig:nxy_ising_Tc}. Here we see that for $L \geq 60$, the pseudo-critical temperatures fit reasonably well on a straight line when plotted against $L^{-1}$. Fitting a line to the largest four lattice sizes yields $$T_c = 0.3315 \pm 0.0002.$$

\begin{figure}[h]
    \centering
    \scalebox{0.44}{\includegraphics{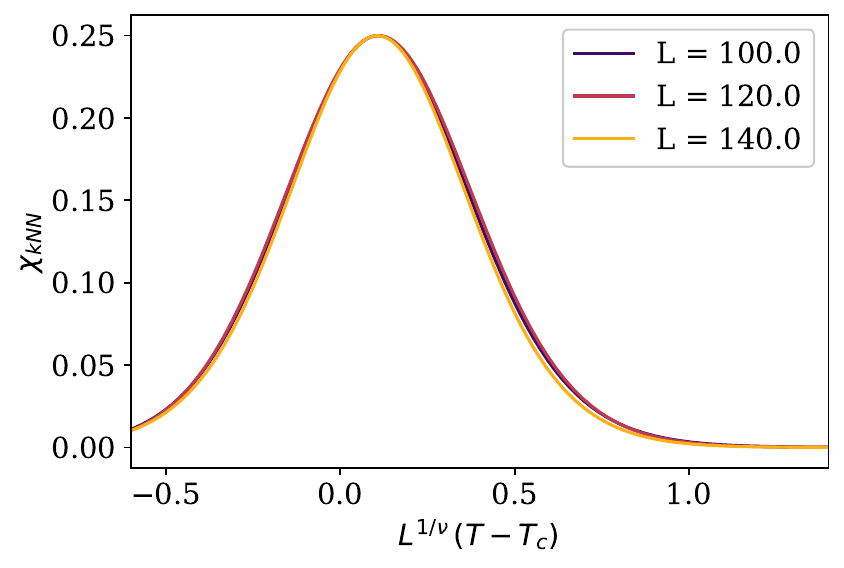}}
    \caption{The curve collapse of $\chi_{kNN}$ for the Magnetic-Nematic transition in the Nematic XY model with $T_c = 0.3316$ and $\nu = 0.9551$.}
    \label{fig:nxy_ising_knn_collapse}
\end{figure}

The curve collapse (Figure \ref{fig:nxy_ising_knn_collapse}) procedure gives
\begin{center}
\begin{tabular}{ c }
 $T_c = 0.3316 \pm 0.0002$ \\ $\nu = 0.9551 \pm 0.0196,$
\end{tabular}
\end{center}
very close to the expected value of $T_c = 0.3314$, but not quite compatible with $\nu = 1$ although better than the logistic regression result.

\begin{figure*}[t]
    \centering
    \scalebox{0.45}{\includegraphics{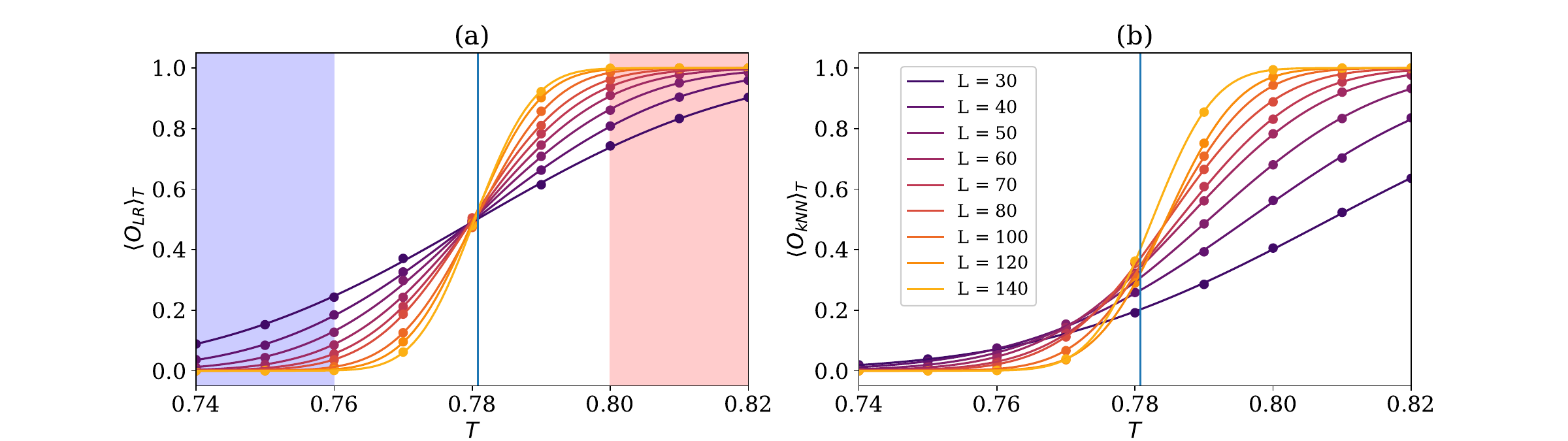}}
    \caption{Plots showing (a) $\langle O_{LR} \rangle$ and (b) $\langle O_{kNN} \rangle$ as a function of temperature for each lattice size for the Nematic-Paramagnetic transition in the Nematic XY model. The shaded regions indicate the temperatures used for the low and high temperature training data. The vertical line shows the location of the expected critical temperature $T_c = 0.7808$. Note that for the k-nearest neighbours case the training regions lie outside the bounds of the plot.}
    \label{fig:nxy_bkt_curves}
\end{figure*}

\subsubsection{Logistic Regression Analysis of Nematic-Paramagnetic Transition}

\begin{figure*}[t]
    \centering
    \scalebox{0.36}{\includegraphics{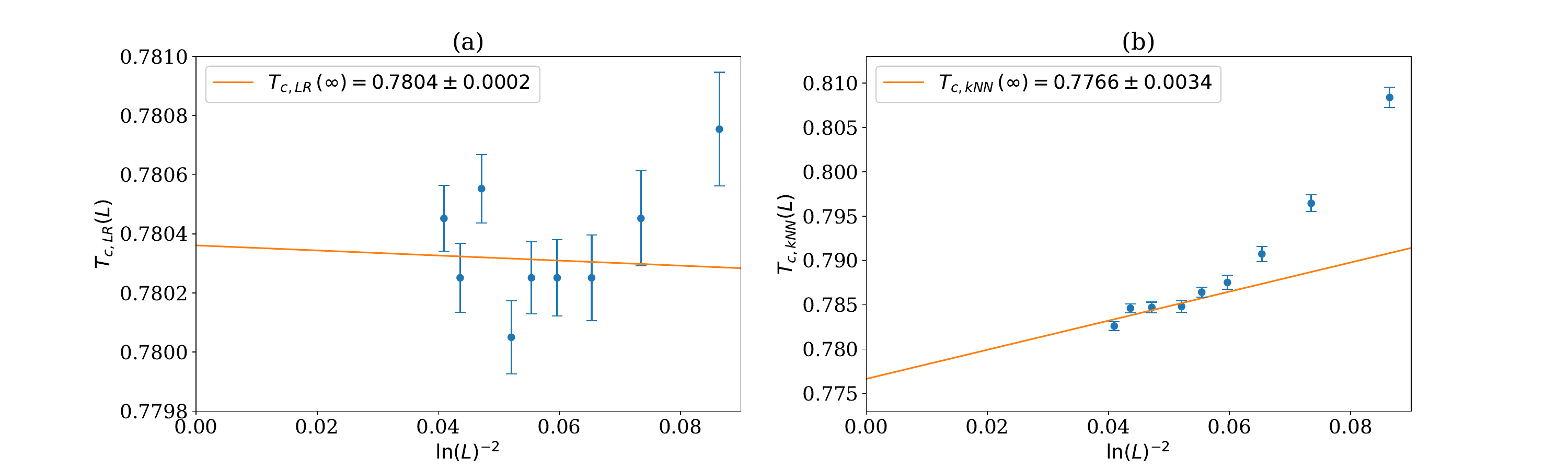}}
    \caption{Estimating the critical temperature for the Nematic-Paramagnetic transition in the Nematic XY model using (a) logistic regression and (b) k-nearest neighbours. The pseudo-critical temperatures for the different lattice sizes, calculated from finding the peak of $\chi_p$, are fitted to the ansatz in Equation \ref{eqn:bkt_tc_scaling}. For the logistic regression we use all the lattice sizes except the smallest in the fit, and for the k-nearest neighbours we use the largest four lattice sizes. The intercept gives the estimate for $T_c(\infty)$. Error bars are estimated by bootstrapping.}
    \label{fig:nxy_bkt_Tc}
\end{figure*}

\begin{figure}[h]
    \centering
    \scalebox{0.44}{\includegraphics{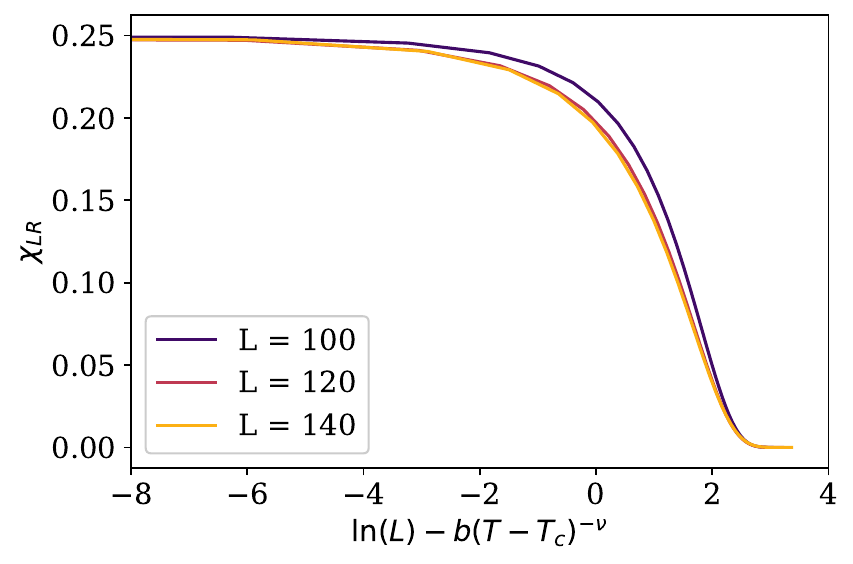}}
    \caption{The curve collapse of $\chi_{LR}$ for the Nematic-Paramagnetic transition in the Nematic XY model with $T_c = 0.7803$, $\nu = 0.5107$ and $b = 0.3037$.}
    \label{fig:nxy_bkt_lr_collapse}
\end{figure}

We trained logistic regression models on samples drawn from $T = 0.74$, $0.75$ and $0.76$ in the low temperature phase, and $T = 0.8$, $0.81$ and $0.82$ in the high temperature phase with $10,000$ samples from each. The regularisation hyper-parameter was set to $C = 0.001$. We evaluated the models with $10,000$ samples from each of $T = 0.74$, $0.75$, \dots, $0.82$. A plot of the resulting phase indicators is shown in Figure \ref{fig:nxy_bkt_curves}. The plot of the pseudo-critical temperatures against $log(L)^{-2}$ is shown in Figure \ref{fig:nxy_bkt_Tc}. We do not observe any significant lattice-size dependence in the pseudo-critical temperatures. They instead seem to be distributed just above to $T = 0.78$ which is the midpoint of the training temperatures. While a straight line fit to all but the smallest lattice size yields an extrapolated critical temperature of $$T_c = 0.7804 \pm 0.0002,$$ not too far from the expected $T_c \approx 0.7808$, Figure \ref{fig:nxy_bkt_Tc} does not suggest that increasing the statistics would lead to increased accuracy. However, the curve collapse (Figure \ref{fig:nxy_bkt_lr_collapse}) procedure gives
\begin{center}
\begin{tabular}{ c }
 $T_c = 0.7803 \pm 0.0025$ \\ $\nu = 0.5107 \pm 0.0101$ \\ $b = 0.3037 \pm 0.0076,$
\end{tabular}
\end{center}
accounting for the expected value of $T_c = 0.7808$, but giving a potentially questionable result for $\nu = \frac{1}{2}$ which lies just outside one standard deviation. 

\begin{figure}[h]
    \centering
    \scalebox{0.5}{\includegraphics{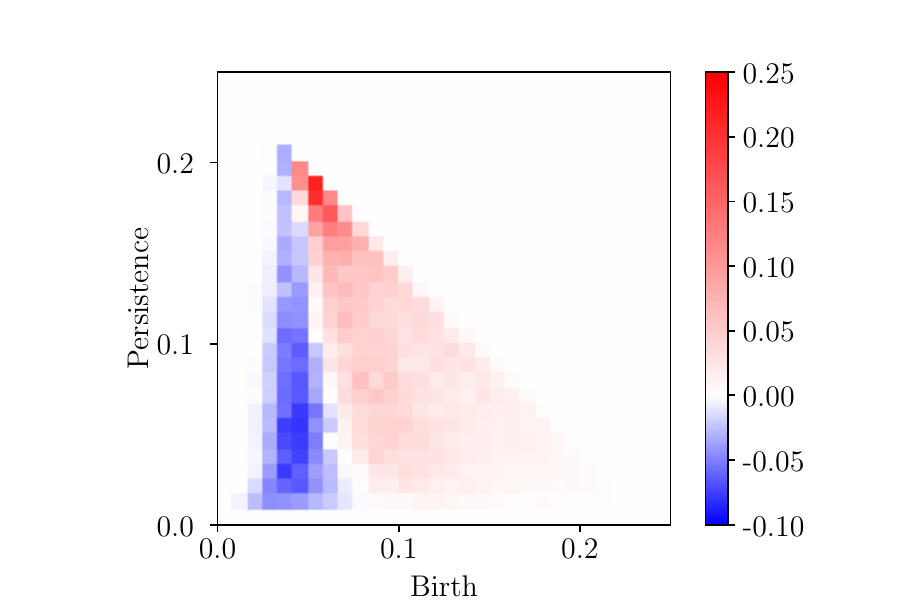}}
    \caption{The weights of the logistic regression model trained on configurations over the Nematic-Paramagnetic transition in the Nematic XY model with $L = 140$.}
    \label{fig:nxy_bkt_coefs}
\end{figure}

The weights of the logistic regression model trained for $L = 140$ are shown in Figure \ref{fig:nxy_bkt_coefs}. We note the similarity to the weights learnt for the XY model in Figure \ref{fig:xy_coefs} except now the region in the top left represents half-vortices and half-antivortices which change behaviour, shifting down to the right as temperature increases and they unbind.

\subsubsection{k-Nearest Neighbours Analysis of Nematic-Paramagnetic Transition}

\begin{figure}[h]
    \centering
    \scalebox{0.44}{\includegraphics{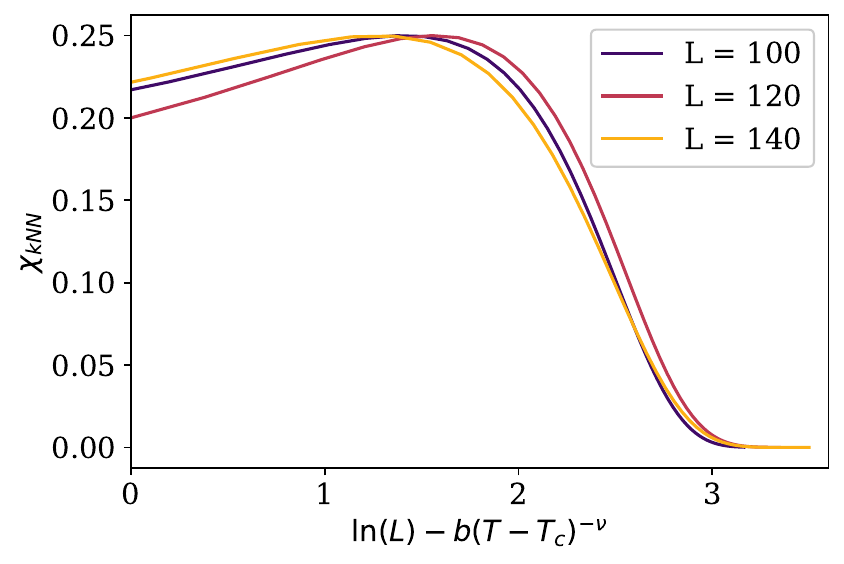}}
    \caption{The curve collapse of $\chi_{kNN}$ for the Nematic-Paramagnetic transition in the Nematic XY model with $T_c = 0.7757$, $\nu = 0.4983$ and $b = 0.3051$.}
    \label{fig:nxy_bkt_knn_collapse}
\end{figure}

Similarly to the case of the XY model, we found that the k-nearest neighbours classification worked best when trained on a broad range of temperatures. We trained the models on samples drawn from $T = 0.5$, $0.55$, \ldots, $0.7$ in the low temperature phase, and $T = 0.85$, $0.9$, \ldots, $1.05$ in the high temperature phase with $2000$ samples from each. The neighbours hyper-parameter was set to $k = 30$. We evaluated the models with $10,000$ samples from each of $T = 0.74$, $0.75$, \ldots, $0.82$. A plot of the resulting phase indicators is shown in Figure \ref{fig:nxy_bkt_curves}. The plot of the pseudo-critical temperatures against $\log(L)^{-2}$ is shown in Figure \ref{fig:nxy_bkt_Tc}. Here we see an asymptotic convergence towards a linear dependence between the pseudo-critical temperatures $T_c(L)$ and $\log(L)^{-2}$. Fitting a straight line to the largest four lattice sizes yields $$T_c = 0.7766 \pm 0.0034.$$ While this is further from the expected $T_c \approx 0.7808$ than the result of the logistic regression approach, the approach towards the correct finite-size scaling is much clearer. The curve collapse (Figure \ref{fig:nxy_bkt_knn_collapse}) procedure gives
\begin{center}
\begin{tabular}{ c }
 $T_c = 0.7757 \pm 0.0064$ \\ $\nu = 0.4983 \pm 0.0226$ \\ $b = 0.3051 \pm 0.0083,$
\end{tabular}
\end{center}
which is compatible with the expected values of $T_c = 0.7808$ and $\nu = \frac{1}{2}$.

\section{Conclusions and Discussion}
\label{sec:conclusions}

We have introduced a new way of applying persistent homology to analyse the configurations of lattice spin models, investigating the phase transitions in the 2D XY model with three different Hamiltonians: the standard action, a topological lattice action, and a modified standard action with an additional nematic interaction term. In each case we were able to successfully identify the phase transition and estimate its critical temperature and critical exponent of the correlation length by considering the finite-size scaling of observables derived from the persistent homology of configurations. In particular we trained logistic regression and k-nearest neighbours classifiers to identify the phases of the models from persistence images. The critical point was estimated as the temperature at which the variance in the classification reached a maximum. 

We have found that the previously introduced approach of using logistic regression for classification, while useful for interpreting which regions of the persistence image indicate the different phases, fails to produce accurate estimates of the critical temperature or exponents in the case of the BKT transitions. Instead it will tend to yield the midpoint between the low and high-temperature training temperatures as the critical temperature. Indeed, using different temperatures for the training causes the estimated critical temperature to shift accordingly. We believe this failure is because logistic regression is a generalised linear model and the data here is highly non-linear. On the other hand, the non-parametric k-nearest neighbours approach generally produces good results, with a clear asymptotic approach towards the expected finite-size scaling behaviour in all cases.

There are a number of interesting questions and directions for further research: 
\begin{itemize}
\item The approach presented in this paper could easily be extended to other lattice spin models, but it would also be interesting to see if the filtration presented in Section \ref{sec:filtration} could be adapted to more complex models such as those from lattice gauge theory.
\item The similarity of the persistence images across the BKT transition in all three models raises the question of the extent to which the persistence is a universal quantity. This could potentially facilitate a transfer learning approach where classifiers trained on one model can identify phase transitions of the same universality class in another model (see, e.g., \cite{Bachtis:2020ajb}).
\item It could also be investigated if the use of a vectorisation and a classifier is necessary in the first place. There is a notion of variance for persistence diagrams called Fr\'echet variance \cite{Turner2014FrchetMF} which might show finite-size scaling behaviour directly. However this is computationally expensive to measure.
\item Finally, we note that there have been a variety of different filtrations used to compute the persistent homology of configurations of lattice spin models. It would be interesting to see how these perform and complement one another on a single data set.
\end{itemize}

\begin{acknowledgments}
Numerical simulations have been performed on the Swansea SUNBIRD system. This system is part of the Supercomputing Wales project, which is part-funded by the European Regional Development Fund (ERDF) via Welsh Government. Persistent homology calculations were performed using giotto-tda \cite{tauzin2020giottotda}. NS has been supported by a Swansea University Research Excellence Scholarship (SURES). JG was supported by EPSRC grant EP/R018472/1. BL received funding from the European Research Council (ERC) under the European Union’s Horizon 2020 research and innovation programme under grant agreement No 813942. The work of BL was further supported in part by the UKRI Science and Technology Facilities Council (STFC) Consolidated Grant ST/T000813/1, by the Royal Society Wolfson Research Merit Award WM170010 and by the Leverhulme Foundation Research Fellowship RF-2020-461{\textbackslash}9. 
\end{acknowledgments}

\appendix

\section{Cubical Complexes and Homology}
\label{appendix:cubical}

This is a very compressed version of the exposition found in \cite{kaczynski2004computational}. An \textit{elementary interval} is an interval of the form $[i, i+1] \subset \reals$ (\textit{non-degenerate}) or $[i,i] = \{n\}$ (\textit{degenerate}) for some choice of $i \in \integers$. An \textit{elementary cube} is a finite product of elementary intervals $Q = I_1 \times \ldots \times I_n \subset \reals^n$, where $n$ is some fixed \textit{embedding dimension}. Its \textit{dimension} $\text{dim}\,Q$ is the number of non-degenerate intervals in the product. A \textit{cubical complex} $C$ is a subset of $\reals^n$ which is a union of elementary cubes. Specifying a field $\mathbf{F}$, we define $\mathbf{F}$-vector spaces $C_k = \{ \sum \alpha_i Q_i \mid Q_i \subseteq C \text{, } \text{dim}\,Q_i = k \text{, } \alpha_i \in \mathbf{F} \}$ for each $k \in \nats$, consisting of finite formal sums of elementary cubes. The \textit{boundary} of a non-degenerate elementary interval is given by the formal sum $\partial [i, i+1] = [i+1,i+1] - [i,i]$. For a degenerate elementary interval the boundary is zero. The boundary of an elementary cube $Q = (I_1 \times \ldots \times I_n)$ is a formal sum 
\begin{equation}
\label{eqn:boundary}
    \partial Q = \sum_{j=1}^n (-1)^{\sum_{i=1}^{j-1}\text{dim } Q_i} (I_1 \times \ldots \times \partial I_j \times \ldots \times I_n)
\end{equation} where we consider $\times$ as distributing over the formal summation. We can see that for $\text{dim}\,Q \geq 1$ we have $\text{dim}\,\partial Q = \text{dim}\,Q - 1$. Therefore we can extend $\partial$ to linear maps $\partial_k : C_k \rightarrow C_{k-1}$ via the mapping $\sum \alpha_i Q_i \mapsto \sum \alpha_i (\partial Q_i)$. Since $\partial \partial I = 0$ for any elementary interval $I$, we also see that $\partial_{k} \circ \partial_{k+1} = 0$ for all $k\in\nats$., so that $\text{im}\,\partial_{k+1} \subseteq \text{ker}\,\partial_{k}$. A sequence of linear maps
$$\ldots \rightarrow C_3 \xrightarrow{\partial_3} C_2 \xrightarrow{\partial_2} C_1 \xrightarrow{\partial_1} C_0 \xrightarrow{\partial_0} 0$$
with this property is called a \textit{chain complex}. The \textit{$k$\textsuperscript{th} cubical homology} of $C$ over $\mathbf{F}$ is defined to be the quotient vector space
$$H_k(C;\mathbf{F}) = \frac{\text{ker}\,\partial_k}{\text{im}\,\partial_{k+1}}.$$
This construction is functorial: given a suitable definition of a \textit{cubical map} $f: C \rightarrow D$ between cubical complexes, there is an induced map $f_k : H_k(C;\mathbf{F}) \rightarrow H_k(D;\mathbf{F})$ for each $k\in\nats$. We will not introduce these in general, but will note that given $C \subseteq D$, the inclusion map $C \xhookrightarrow{} D$ is cubical and hence induces maps on homology.

\begin{figure}
\centering
\includegraphics{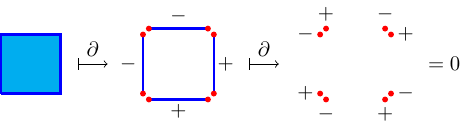}
\caption{Example of how the boundary operator $\partial$ acts on a simple cubical complex consisting of a single $2$-dimensional cube. Note how sum in equation \eqref{eqn:boundary} being alternating ensures that $\partial\partial = 0$.}
    \label{fig:cubical_boundary}
\end{figure}

\section{Stability of Filtration}
\label{appendix:stability}

We briefly show that the filtrations introduced in Section \ref{sec:filtration} yields persistent homology which is stable under small perturbations of spins with respect to the bottleneck distance between persistence diagrams. We make use of the following concept:
\begin{definition}[Interleaving]
    Given filtrations $$F,G : (\reals, \leq) \rightarrow \mathbf{CubicalComplex},$$we say that $F$ and $G$ are $\epsilon$-interleaved if for all $t \in \reals$ we have that $F(t) \subseteq G(t+\epsilon)$ and $G(t) \subseteq F(t+\epsilon)$. We say that the interleaving distance between $F$ and $G$ is $$d_{I}(F, G) = inf \{\epsilon \,\vert\, \text{$F$ and $G$ are $\epsilon$-interleaved}\}.$$
\end{definition}
Then from \cite{Bubenik2014CategorificationOP} we have the following theorem.
\begin{theorem}[\cite{Bubenik2014CategorificationOP} Proposition 3.6 and Theorem 4.16]
    Given filtrations $F$ and $G$ as before, and $k \in \nats$, we have that
    $$d_B(PH_k(F), PH_k(G)) \leq d_I(F, G)$$where $d_B$ is the bottleneck distance.
\end{theorem}
 Thus we just need to show that the filtration we assign to a configuration $F_{\boldsymbol{\theta}}$ are $\epsilon$-interleaved with the filtration obtained after a perturbation of the configuration for some $\epsilon$ bounded by the changes in the spins. Take $F$ to be the angle difference filtration introduced in Section \ref{sec:filtration} and suppose we have a configuration $\boldsymbol{\theta}$ and change spin $\theta_i$ to $\theta_i + \Delta \theta$ to obtain a configuration $\boldsymbol{\theta}^\prime$. Then given a neighbouring spin $\theta_j$, denote the length of the smallest arc between $\theta_i$ and $\theta_j$ by $d_{ij}$ and the length of the arc between $\theta_i + \Delta\theta$ and $\theta_j$ by $d_{ij}^\prime$. Then we have that $$d_{ij} - \Delta\theta \leq d_{ij}^\prime \leq d_{ij} + \Delta\theta.$$
Hence for all filtration values $t\in\reals$ we have inclusions $F_{\boldsymbol{\theta}}(t) \hookrightarrow F_{\boldsymbol{\theta}^\prime}(t+\Delta\theta / 2\pi)$ and $F_{\boldsymbol{\theta}^\prime}(t) \hookrightarrow F_{\boldsymbol{\theta}}(t+\Delta\theta / 2\pi)$ so that the filtrations are $\frac{\Delta\theta}{2\pi}$-interleaved. By the theorem above, we have $$d_B(PH_k(F_{\boldsymbol{\theta}}), PH_k(F_{\boldsymbol{\theta}^\prime})) \leq \frac{\Delta\theta}{2\pi}$$ for all $k\in\nats$. A straightforward application of the triangle inequality shows that if multiple spins are altered then the bottleneck distance is bounded by the sum of the alterations. The same argument applies in the case of the nematic angle difference filtration.

On the other hand, we observe that the sublevel set filtration for $\mathbb{S}^1$-valued spins introduced in \cite{cole2020quantitative} does not have this stability. An arbitrarily small perturbation $\epsilon$ to one of the spins can change its angle from $\pi$ to $-\pi + \epsilon$, potentially introducing or removing a high persistence point in the persistence diagram. Although this is mitigated to an extent in practise by choosing angle $0$ to be in the direction of the magnetisation.

\section{Histogram Reweighting}
\label{appendix:histogram_reweighting}
Histogram reweighting allows us to express the ensemble average of an observable $O$ at temperature $T^\prime$ in terms of averages at any other temperature $T$ according to the equation
\begin{equation}
\label{eqn:single_histogram_reweighting}
    \langle O \rangle_{T^\prime} = \frac{\langle O e^{-(\beta^\prime - \beta) E} \rangle_T}{\langle e^{-(\beta^\prime - \beta) E} \rangle_T}
\end{equation}
where $\beta = 1/T$, $\beta^\prime = 1/T^\prime$, and $E$ is the energy of a configuration \cite{PhysRevLett.61.2635}. However, in practice we can only reweight so far, so that the energy distributions for $T$ and $T^\prime$ have a sizable overlap. To reliably extrapolate to a wider region we can make use of multiple histogram reweighing \cite{PhysRevLett.63.1195} where we sample at multiple temperatures $T_1, \ldots , T_R$ (with corresponding inverses $\beta_1, \ldots , \beta_R$). Suppose we sample $N_i$ configurations at temperature $T_i$, then we can iterate the equation
$$e^{-f_{\beta}} = \sum_{i=1}^R \sum_{a=1}^{N_i} \frac{g_i^{-1}e^{-\beta E_i^a}}{\sum_{j=1}^R N_j g_j^{-1} e^{-\beta_j E_i^a + f_j}}$$
to estimate the free energies $f_{i} = f_{\beta_i}$ at the temperatures $T_i$ up to an additive constant, where each $g_i$ is a quantity related to the integrated autocorrelation of the samples in run $i$. Given the $f_i$ we can estimate
$$\langle O \rangle_{T^\prime} = \sum_{i=1}^R \sum_{a=1}^{N_i} \frac{O_i^a g_i^{-1}e^{-\beta_k E_i^a + f_{\beta^\prime}}}{\sum_{j=1}^R N_j g_j^{-1} e^{-\beta_j E_i^a + f_j}}.$$

\section{Bootstrap Error Estimation}
\label{appendix:bootstrap}
In order to make any reasonable conclusions from the results of our analysis we need to be able to estimate the error in any numerical values obtained. While the error in ensemble averages can be directly estimated from the sample, we also calculate various fits to the data. The way in which error propagates here is not necessarily easy to calculate directly. Recall that the idea of bootstrap analysis is to sidestep these concerns by estimating the sampling distribution of a statistic directly. Suppose we obtain $N$ sampled configurations $S = \{ \boldsymbol\theta_1 , \ldots , \boldsymbol\theta_N \}$ and calculate some numerical statistic $f(S)$ from the data. Given some preset integer $N_B$, bootstrap analysis proceeds by:
\begin{enumerate}
    \item resampling $S$ with replacement $N_B$ times to obtain samples $S_1 ,\ldots, S_{N_B}$ each of size $N$; then
    \item computing $f(S_i)$ for each $i \in \{1,\ldots,N_B\}.$
\end{enumerate}
For large enough $N_B$, the distribution of the $f(S_i)$ approximates the sampling distribution of $f$ and we can estimate the standard error $$\sigma_f \approx \sqrt{\frac{1}{N_B - 1}\sum_i\big(f(S_i) - \overline{f(S_j)}\big)^2}.$$


\bibliography{main}

\end{document}